**Advanced Materials and Device Architectures for Magnetooptical Spatial Light Modulators**

*Soheila Kharratian*, Hakan Urey, Mehmet C. Onbaşlı**

S. Kharratian, Dr. M. C. Onbaşlı
Department of Materials Science and Engineering, Koç University, Sarıyer, Istanbul, 34450, Turkey
E-mail: skharratian15@ku.edu.tr, monbasli@ku.edu.tr

Prof. H. Urey
Department of Electrical and Electronics Engineering, Koç University, Sarıyer, Istanbul, 34450, Turkey

Keywords: magnetooptics, spatial light modulators, garnets and magnetooptical oxides, photonic crystals

Faraday and Kerr rotations are magnetooptical (MO) effects used for rotating the polarization of light in transmission and reflection from a magnetized medium, respectively. MO effects combined with intrinsically fast magnetization reversal, which can go down to a few tens of femtoseconds or less, can be applied in magnetooptical spatial light modulators (MOSLMs) promising for nonvolatile, ultrafast, and high-resolution spatial modulation of light. With the recent progress in low-power switching of magnetic and MO materials, MOSLMs may lead to major breakthroughs and benefit beyond state-of-the-art holography, data storage, optical communications, heads-up displays, virtual and augmented reality devices, and solid-state light detection and ranging (LIDAR). In this study, the recent developments in the growth, processing, and engineering of advanced materials with high MO figures of merit for practical MOSLM devices are reviewed. The challenges with MOSLM functionalities including the intrinsic weakness of MO effect and large power requirement for switching are assessed. The suggested solutions are evaluated, different driving systems are investigated, and resulting device architectures are benchmarked. Finally, the research opportunities on MOSLMs for achieving integrated, high-contrast, and low-power devices are presented.

## 1. Introduction

Magnetooptical (MO) effects are the phenomena which result from angular momentum transfer between photons and magnetic moments in a magnetized matter. In the classical picture, light with linear polarization is the superposition of a left- and a right-hand circularly polarized (LHCP and RHCP) light beams with identical amplitudes and a phase difference. The phase difference between LHCP and RHCP components of a linearly polarized light defines the angle of its polarization plane. When light is shone on a material, the angular momenta associated with its circularly polarized components affect the charged particles in the material and result in their circular motions. These motions lead to effective magnetic fields that are in opposite directions for LHCP and RHCP beams. When an external magnetic field is applied, the net field magnitudes experienced by LHCP and RHCP components are different, which causes dissimilar interaction and propagation velocities for each [1],[2]. In a macroscopic point of view, the permittivity values for LHCP and RHCP light would be different. This difference leads to a relative phase accumulation between two components of linearly polarized light and results in polarization rotation of light when it passes through or reflects back from a magnetized medium, named Faraday and Kerr effect, respectively [3],[4],[5]. In quantum mechanical description, magnetooptical effects are generally a second-order perturbation on the combined electron and spin wavefunctions. The incident photon's angular momentum is transferred to the electron's both orbital and spin angular momenta. As a result, photon angular momentum and light polarization are slightly shifted due to the MO effects.

The spectral and composition dependence of MO effects and their intensities provide characteristic signatures on the electromagnetic (EM) waves or electronic and spin structure of materials [6] This makes them suitable for various analytical chemical methods such as visible or near-infrared magnetooptical spectroscopy [7],[8],[9] x-ray magnetic circular dichroism (XMCD) [10],[11],[12] Brillouin light spectroscopy (BLS) [13] in addition to applications such as optical isolators [14],[15],[16] , circulators [17],[18],[19] , spatial light modulators [20],[21],[22], polarized microscopy [23],[24],[25], sensing/imaging systems [26],[27],[28], data storage [29],[30],[31],[32] and growing field of spintronics [33],[34],[35].

Among the above-mentioned applications, spatial light modulators (SLMs) have drawn significant attention since they are the key components of many photonic devices including holograms and display systems, optical interconnects, projectors, functional Raman microscopy, and visible light communications [36],[37],[38],[39],[40]. SLM is an optical device with an array of pixels which use external control signals to modify the amplitude, phase or polarization of a wave front as a function of position [41]. SLMs typically work in phase-only [42],[43],[44],[45] , amplitude-only [46],[47] , or phase-amplitude modes [48],[49] , and can perform binary or analog modulation. Well-developed types of SLMs are digital micromirror devices (DMDs) and liquid crystal (LC)-based devices including liquid crystal display (LCD) and liquid crystal on silicon (LCOS). Table 1 shows different state-of-the-art SLM products meeting different application requirements. Smaller pixel sizes are achieved with LCOS devices (min. pixel pitch of 3.74 µm), but their response times are long (in the order of milliseconds), while produced DMDs have larger pixels (min. pitch of 5.4 µm) with shorter response times (~10 µs).

These SLMs are final products, which cannot yet simultaneously achieve fast modulation and high spatial resolution needed for holography and 3D imaging. For high quality video holograms, SLMs with high space bandwidth product (SBP) are required. SBP is the product of physical size of SLM and the spatial bandwidth, which is determined by the total pixel count in SLM. For a constant SBP, there is a trade-off between the size of viewing window (eye box) and field of view (FoV). For holographic systems with both high FoV and large eye box, SLMs with high pixel counts are needed. Increasing the size of SLM for this purpose is not feasible in most applications due to prohibitively large system sizes, ergonomic considerations, driving optics and electronics. As a result, the solutions focused on reducing the pixel sizes. Large pixels also cause narrow viewing zone and aliasing noise (overlap of diffraction orders) according to the following relation between viewing angle and pixel pitch:

$$\theta = 2\sin^{-1}\left(\frac{\lambda}{2p}\right) \quad (1)$$

where λ is the wavelength and p is the pixel pitch. For an acceptable viewing angle, the pixel pitch should be on the order of the wavelength of light (pixels < 1 µm for visible) [50],[51],[52],[53]. Thus, SLMs with small pixel size, high pixel count, and sufficiently fast to address plenty of pixels in a single image frame, are desired for holography. Research is in progress to reduce the pixel size [45],[54],[55],[56],[57],[58],[59] and response time [60],[61],[62],[63]. Many designs and prototypes have been reported, yet a practical active SLM with large number of small pixels and short response time has not been demonstrated.

SLMs that use MO effects for modulating light, called magnetooptical spatial light modulators (MOSLMs), hold promise to address the above-mentioned challenges associated with current SLMs. In MOSLMs, the polarization plane of light is rotated by MO effects and using polarizers allows for passing only the light components at specific polarization angles. This makes possible the analog or digital modulation of light by controlling the amount of rotation in polarization plane, i.e. the intensity of occurring MO effects. Each pixel can be controlled and switched by manipulating its magnetization state. By reversing the magnetic moments in a pixel, the direction of Faraday/Kerr rotation can be reversed and this phenomenon is observed in the intensity and phase of the out coming beam [64].

Table 2 summarizes some prototype demonstrations of the MOSLMs. Although these devices are not as developed in comparison to the mature products listed in Table 1, there are prominent features unique to MOSLMs that keep the field active, and make them promising for many applications. One of these features is the intrinsic high speed of MO phenomena and magnetization switching, which happen in time scales ranging from nanoseconds [65],[66],[67] down to femtoseconds [68] or less [69].This makes MOSLMs distinctive candidates for ultrafast spatial modulation of light, which can extend the modulation frequency to multi-THz rates. Another advantageous feature accompanying MOSLMs is the magnetic remanence, which makes them nonvolatile devices that can operate as memory elements and save the written data in absence of power. Furthermore, MOSLM is a robust solid-state device that can eliminate disadvantages of mechanically moving parts and complicated fabrication steps in DMDs, or pixel crosstalk due to the fluid shape of material and lack of physical separation between pixels in LC-based devices [37],[38]. Moreover, MOSLMs might allow for modulation with monolithically integrated thin films and reduced pixel sizes. Dynamic control of magnetization switching by current or voltage [70],[71],[72], turns MOSLMs into integrated active devices. Combination of these features makes MOSLMs outstanding candidates for holographic applications and 3D displays [73],[74],[75], augmented (AR) and virtual reality (VR) devices, LIDAR, beam steering devices for photonic projectors and other imaging applications [76][77], optical isolators and circulators [78], and visible light communication [79].

Comparing Tables 1 and 2, early MOSLM prototypes could reach frame rates above 1 kHz for visible which are more advantageous than state-of-the-art SLMs. Power consumption in SLM products are within 1-50 W while MOSLM prototype switching powers are less than 1 W. Power requirements reported for MOSLM prototypes do not include complicated driving electronics and optics, so a final power comparison between mature SLM types and MOSLMs should be done after MOSLMs turn into products. On the other hand, the pixel pitch sizes, pixel counts, wavelength ranges (visible only) and the modulation types (binary only) in MOSLMs must be improved by using the recent MO materials breakthroughs and new switching physics. The main issue hindering MOSLM from practical device applications is the inherent weakness of MO effects. The modulation depth and pixel contrast in MOSLMs depend on the Faraday and Kerr rotation angles. The magnitude of Faraday rotation (FR) is proportional to the path length of light in the MO material while Kerr rotation is generally not large enough (~milliradians) for practical device applications. Hence, enhanced MO effects and reduced optical loss are essential while miniaturizing and integrating MO devices [1],[22],[80]. Another important challenge with MOSLMs is their high power consumption. This problem originates from the large magnetic fields required for switching pixels, the optical source power needed to offset the high optical losses in MO materials and Joule heating. Joule heating leads to switching errors and reliability issues [51]. Therefore, MOSLMs have been held back due to low figure-of-merit materials, inefficient switching mechanisms and the associated device architectures.

Researchers recently demonstrated major breakthroughs in materials and underlying physics of magnetooptics and magnetization switching mechanisms. Advanced synthesis and growth techniques (pulsed laser deposition and sputtering) enable high-quality MO materials fabrication. Precise micro/nanofabrication and characterization methods allow for targeting small pixel sizes. The recently discovered mechanisms of low power control of magnetism may help significantly reduce power consumption. These developments offer significant untapped potential that could enable in the near future a new generation of nonvolatile, ultrafast and low-power MOSLMs.

In this review, we evaluate the recent progress in magnetooptical materials and their applications for spatial light modulation. We then discuss the challenges and perspectives for MO devices. Recently, different reviews [81],[82],[83],[84] have been published on MO garnets and their applications in photonic integrated circuits (PICs). However, there are no comprehensive reviews linking the recently developed high figure-of-merit MO materials and low power spintronic switching mechanisms to high-contrast and compact MOSLMs. This review aims to fill this gap and provide an evaluation of the research opportunities to guide the field towards a new generation of practical integrated MOSLMs.

In section 2, we discuss different MO materials as potential constituents of MOSLMs and report the progress to improve the practicality and efficiency of MOSLMs. In section 3, we review the methods for enhancing MO effects with a focus on photonic crystal and plasmonics. In section 4, we present the conventional and the recently developed low power spintronic switching methods for MOSLMs. In section 5, we attempt at bridging the progress covered in sections 2-4 towards nonvolatile, ultrafast and low power MOSLMs by covering the desired materials/device specifications and suggesting implementations for spanning different spatial and temporal resolutions of modulation.

## 2. Magnetooptical materials

Magnetooptical effects mostly occur in magnetic materials with nonzero magnetization such as ferromagnets and ferrimagnets; however, these phenomena have been observed in antiferromagnetic materials as well [85],[86],[87] .

The property of a magnetooptical or gyrotropic material with a magnetization along z direction can be described with an antisymmetric permittivity tensor in the form of

$$\varepsilon = \begin{pmatrix} \varepsilon_1 & +i\varepsilon_2 & 0 \\ -i\varepsilon_2 & \varepsilon_1 & 0 \\ 0 & 0 & \varepsilon_3 \end{pmatrix} \qquad (2)$$

where all tensor elements are complex numbers having real and imaginary parts. Below Curie temperature $T_c$ the value of $\varepsilon_3$ is very close to $\varepsilon_1$ and above $T_c$ it is equal to $\varepsilon_1$. With a restriction of low magnetic field and sticking to linear MO effects, it can be safely assumed $\varepsilon_3 = \varepsilon_1$ which simplifies the matrix to two, diagonal and off-diagonal, elements [88],[89],[90] . $\varepsilon_1$ is related to the regular refractive index n and extinction coefficient k. The off-diagonal elements are related to refractive indices $n_\pm$ and extinction coefficients $k_\pm$ of LHCP and RHCP light, and the magnitude of $\varepsilon_2$ indicates the strength of the MO response [91],[92] . The LHCP and RHCP components of a linearly polarized light, undergoing different refractive indices, propagate with different speeds, and this builds up a relative phase difference between them. In this conditions, the polarization plane of the resultant linearly polarized wave rotates (Faraday/Kerr rotation). Moreover, the difference in the extinction coefficients of LHCP and RHCP components causes different absorptions and consequently, different amplitudes at the output, which yields Faraday/Kerr ellipticity. While the diagonal elements of the permittivity tensor do not directly depend on magnetization (M), $\varepsilon_2$ grows mainly linearly with M [93].

Materials that exhibit MO effects can be distinguished in three different categories. In the first category, MO effects originate from the direct action of magnetic field on the orbital motion of electrons and $\varepsilon_2$ is essentially a function of magnetic field (H). All diamagnetic materials including organic molecules and planar molecules with at least uniaxial symmetry fall into this category. In the second category, spin-orbit coupling of aligned spins is the main motive of MO effects, while the direct impact of magnetic field on the electronic orbital motion is negligible. In other words, magnetic interaction of an oriented spin has a much stronger effect on the orbital motion of an electron compared to the direct effect of an external magnetic field. Ferromagnets and nonmetallic paramagnets at low temperatures are in this category, and it is more appropriate to indicate $\varepsilon_2$ as a function of magnetization (M) instead of H in this group of materials. Semiconductors and nonferromagnetic metals establish the third category which represents a transition between these two extremes. In this category, both orbital motion and spin-orbit interactions can have a noticeable role in the MO phenomena and there is no explicit distinction [94] . Misemer [95] quantitatively investigated the influence of spin-orbit interactions on the MO effects, and indicated that there is an approximately linear relation between strength of spin-orbit coupling and MO coefficients in transition metals.

Magnetooptics involves a close interplay between absorption and polarization rotation, which both strongly depend on the electronic band structure. One could study various MO materials under dielectrics, metals, semimetals and 2D materials categories. Below, we present the materials falling in each class and review the materials engineering efforts for improving the performance of MO materials. Due to their high reflectivity, metals are mainly used in reflection configuration and for Kerr effect; while Faraday effect, mostly achievable in wide bandgap magnetic dielectrics, is more useful for practical device purposes [1].

### 2.1. Dielectrics

Dielectric MO materials include different magnetic oxides, ferrites, spinels, sulfides, and trihalides [1] . Ferrimagnetic garnets are the most important class of MO materials that have been studied extensively during the last decades. A summary of the MO materials synthesized by various methods with their corresponding MO properties is presented in Table 3. Since high FR and low optical loss are both essential in realizing high-contrast MOSLMs with low power consumption, the MO materials figure of merit is:

$$\text{FoM}\ (\text{deg}\cdot\text{dB}^{-1}) = \frac{\text{Faraday rotation}\ (°)}{\text{Optical loss}\ (\text{dB})} \tag{3}$$

and we calculated the best FoM achieved in the literature for fundamental red, green, and blue wavelengths, using the provided data and finite-difference time-domain (FDTD) simulations. From Table 3, one could conclude that in most cases, FoM is smaller than 1 $°\cdot dB^{-1}$, however, substitution of elements like Bi in yttrium iron garnet ($Y_3Fe_5O_{12}$) as a common MO material can help improve FoM significantly [96]. Iron borate ($FeBO_3$) has exceptionally high MO FoM but it has been studied less intensively for MO applications. Difficulty in growing large single crystals of this material is one obstacle for its use in MO devices. Birefringence of $FeBO_3$ also complicates its study [97],[98].

### 2.2. Metals

In metals and alloys, MO properties depend on the density difference of spin-up and spin-down electrons near the Fermi level and the oscillator strength of the optical transitions. Due to such a dependence, it is not intuitive to directly link MO effects with their microscopic origin in the metallic systems, nonetheless, first-principles calculations can provide hints [1].

Magnetic transition metals such as Fe, Ni, Co and their alloys, rare earth-transition metal compounds and intermetallics exhibit MO effects and potentially can be used in MOSLMs;

however, their practical application for SLM devices is generally complicated due to small magnitude of the effects or large absorption in metals [99],[100]. MO Kerr effect (MOKE) in metals was used for measuring magnetic hysteresis loops, for imaging magnetic domains, and studying magnetic dynamics with high resolution [101],[102],[103],[104]. **Figure 1** shows polar MOKE values as a function of magnetization for some metallic compounds in comparison with some insulators. Polar MO Kerr rotation angle $|\theta_K|$ was measured for various ferro-, ferri-, and antiferromagnetic materials at room temperature. In most ferri/ferromagnetic films, $|\theta_K|$ increases proportionally with magnetization, i.e. $|\theta_K| = K_sM$, with $K_s$ being a coefficient within 0.2-2 deg.$T^{-1}$ (the shaded region in Figure 1). $Mn_3Sn$, as an antiferromagnetic metal, has a large MOKE with $K_s$ = 25.6 deg.$T^{-1}$ while for antiferromagnetic insulators, $K_s$ has a value in the range of 10-20 deg.$T^{-1}$ [87].

MO activity has been reported even in the noble metals, but with a much weaker intensity compared to the ferromagnets [105].

### 2.3. Semimetals and 2D materials

Although MO devices are normally based on the conventional MO materials which were discussed above, lately remarkable MO properties were demonstrated in semimetals and two-dimensional (2D) materials which promise for groundbreaking high-density MO and spintronic devices.

Graphene, being the first prototype of 2D materials, has a semimetallic nature with unique mechanical, optical and thermal properties [106] . This atomically thin layer of graphite has been extensively studied for its MO properties in the recent years. After theoretical research reports on MO effects in graphene [107],[108] , in 2011 Crassee et al. [109] for the first time experimentally investigated MO properties of graphene epitaxially grown on a SiC substrate. They observed increasing Faraday rotation (FR) with increasing magnetic field and reported rotations as large as 0.1 rad (~6°) for a single layer graphene in 7 T magnetic field at 5 K (**Figure 2a**). They also discovered a strong magnetic field dependence for transmission and absorption in graphene, as indicated in **Figure 2b**.

It was shown that the giant FR in graphene could be further enhanced by constructive Fabry-Perot interference from substrate, and simulation results presented Faraday rotations up to 0.15 rad (~9°) in multilayer epitaxial graphene grown on SiC [110]. Shimano et al. [111] using THz time-domain spectroscopy (THz-TDS) observed not only Faraday but also Kerr rotation in graphene. Falkovsky [112] explained these effects by appearance of a Hall component in the conductivity tensor of graphene under applied magnetic field, which breaks rotational symmetry around the major axis and implies polarization rotation for a linearly polarized EM wave. He showed that in a free suspended graphene in presence of a 7 T magnetic field, FR as high as 0.25 rad is achievable. Manipulation and tuning of MO properties of graphene with applied strain [113],[114] or electrostatic doping [115] at zero or fixed magnetic fields are other advancements in the field that introduce new ways to control MO effects in novel opto-electro-mechanical devices.

In addition to pristine graphene, MO effects have been sought in nitrogen-graphene crystals, which are graphene with different substitutions of C atoms with N. Based on first-principles calculations, in N-C 2D materials, Faraday and Kerr rotation angles strongly depend on carrier carrier concentration and can be tuned by gate voltage [106]. Silicene, a 2D allotrope of silicon, is also capable of outstanding MO properties. It has been theoretically demonstrated that a silicene monolayer exhibits maximum of 8° and 13° respectively for Faraday and Kerr rotations in terahertz regime [116]. Other 2D materials with MO properties that have been studied so far include phosphorene [117], molybdenum disulphide ($MoS_2$) [118], tungsten diselenide ($WSe_2$) [119], chromium triiodide ($CrI_3$) [120],[121],[122], and $Cr_2Ge_2Te_6$ [123],[124].

More recently, MO effects have been demonstrated in 3D and bulk semimetals. MO properties of Weyl semimetals [125],[126],[127] introduce a way to discriminate them from Dirac semimetals. MO measurements can help probe chiral anomaly in Weyl semimetal state, which can be hosted in noncentrosymmetric and nonmagnetic monoarsenides/phosphides of transition metals, such as TaAs [128] and NbP [129]. Zhang et al. [130] reported a different behavior in $Cd_3As_2$ bulk single crystal which is recognized as a Dirac semimetal. Performing rotational magnetooptical Kerr effect measurements, they showed that in presence of only magnetic field, no Kerr effect is observed. However, applying a current across the sample alongside the magnetic field results in a Kerr rotation angle, which is maximum when the magnetic and electric fields are parallel. The magnitude of Kerr angle increased with increasing magnetic field or current density.

While these materials provide new mechanisms and giant MO figures of merit, their functionalities under ambient conditions and with small magnetic fields are yet to be demonstrated for integrating with MOSLMs.

### 2.4. Materials engineering for improving MO properties

Various materials with different stoichiometries and structures, synthesis methods and parameters, as well as fabrication of composite MO materials were studied in order to achieve larger MO effects. Fabrication procedures and parameters have important effects on MO properties since they determine the final stoichiometry of the materials [131] . Post-deposition treatments and annealing at proper temperatures and durations are used to obtain desired phases and improve crystallization for better MO properties [132],[133],[134],[135].

Among numerous efforts for achieving larger rotation angles in nonreciprocal devices, doping iron garnets with Bi [136],[137],[138],[139] ,Ce [140],[141],[142],[143], or Nd [144] has become a standard method. Doped iron garnets have much higher figures of merit with respect to undoped garnet and can thus be used as typical MO components of these devices. Hansen et al. [145],[146] demonstrated a linear dependence of FR on Bi content in Bi-substituted garnets.

Although garnets were epitaxially grown on lattice-matched Gadolinium Gallium Garnet (GGG) substrates [147], Sung et al. [148] made a breakthrough in 2005, by growing yttrium iron garnet (YIG) on nongarnet substrates. They used RF sputtering and *ex situ* rapid thermal annealing as a fast and reliable fabrication process. However, similar attempts for integrating doped garnets to nongarnet substrates were not entirely successful, due to thermal expansion and lattice coefficient mismatch, cracking during annealing, formation of secondary phases, or incomplete crystallization [81]. These issues diminish the transmission and MO properties, and need to be resolved. Having a hematite underlayer with spinel structure was reported to be helpful in deposition of Bi-substituted garnets on quartz substrate and considerably improved MO properties of the garnet film [149]. Subsequently, undoped YIG seed layer at the bottom or top was introduced to facilitate crystallization of doped garnets grown on nongarnet substrates and alleviate the mentioned problems [78],[150],[151]. It was later shown that terbium iron garnet (TIG) family do not need a seed layer to grow on nongarnet substrates, even when doped with Bi or Ce [139],[143].

Some other efforts in the context of engineering MO materials include following works: Nur-E-Alam et al. [152] made heterostructures by sandwiching MO films having in-plane magnetization between out-of-plane MO films and obtained high MO quality with near-perpendicular magnetization and low coercivity. Chen et al. [153] reported phosphorus-based glass containing YIG crystals fabricated by the incorporation process and observed that FR of the samples increases as a function of YIG content and decreases by annealing temperature. Sadatgol et al. [154] proposed enhancing the FR using MO metamaterials where nonmagnetic conductor wires are embedded into MO media. They explained that plasmonic resonances are not the origin of this enhancement; but it is generated near the dilute plasma frequency and is tunable by modifying geometry of the embedded metamaterial structure. Enhanced MO Kerr effect in Fe/insulator interfaces was calculated by Gu et al. [155], in proportion to the ratio of $\sigma_{xy}/\sigma_{xx}$ (respectively off-diagonal/diagonal elements of optical conductivity tensor). This enhancement was explained by increased orbital magnetic moments and spin-orbit correlations for the interfacial Fe atoms.

## 3. Enhancement of MO effects and different structures for MOSLMs

As mentioned in the introduction section, for realization of practical miniaturized and integrated devices, MO effects need to be enhanced while retaining optical losses in an acceptable level. Other than materials engineering that was discussed in the previous section, device engineering can serve favorably for this purpose. Efforts in this regard have led to different MOSLM device structures that will be described in the present section.

The first MOSLM was composed of magnetic garnet pixels on a nonmagnetic substrate, operating in transmission mode and being thermally switched by a laser beam [156]. In addition to improvement of resolution, switching sensitivity, and frame rate, Cho et al. [65] demonstrated a reflection-mode MOSLM in which a reflector is used at the back of the MO layer. The light entering from the transparent substrate and traversing the MO film, reflects back from the reflector and passes through the MO film for the second time, and ultimately exits from the substrate (**Figure 3**). The nonreciprocal MO Faraday effect causes a double rotation angle compared to the case without reflector. This was the first device engineering attempt for increasing MO rotation angle and became a part of the device designs later.

Other approaches that have been studied extensively in the literature for enhancing MO effects in active nonreciprocal devices, are identified in the following subsections:

### 3.1. Magnetophotonic crystals

Magnetophotonic crystal (MPC) is one of the concepts studied extensively in the literature, for enhancing MO effects in thin films [25],[76],[157],[158], [159],[160],[161],[162],[163]. MPC is a 1D photonic crystal where MO films are sandwiched as defect layers between two Bragg mirrors consisting of alternating high- and low-index dielectric layers. Breaking continuous translational symmetry by the periodic dielectric layers in a MPC results in a photonic band gap in its optical response. Inclusion of the dielectric defects further breaks such a discrete translational symmetry and leads to appearance of transmission peaks within the band gap. In addition, the MO characteristic dielectric defects, i.e. non-zero off-diagonal element of permittivity tensor, breaks the time-reversal symmetry [164]. When the phase matching conditions are met for a certain wavevector, the MO defects act as optical cavities for the photonic crystal in that wavelength. The resultant cavity modes enhance the optical path length of light in the MO layers and accumulate the MO effects with each pass. **Figure 4** schematically shows a single-defect MPC in the form of substrate/$(H/L)^3/D/(L/H)^3$ (H and L are high and low-index dielectrics, respectively, and D is the defect layer).

Increasing the mirrors layer counts results in better localization of light in the defect layers and improves the quality factor of the cavity. Higher quality factor enhances MO effects. Using more MO defect layers results in stronger MO effects as well. A caveat is that the increased photonic path length in the cavity accumulates optical losses. Therefore, optimizing the number of defect and mirror layers based on the required figures of an application is necessary. The optical and MO response of the MPC are also dependent on other parameters such as the material properties of the constituent layers, their thicknesses and relative positions. Hence, the optimization of these parameters and configuration of the MPC is also essential for achieving high MO rotation with minimum optical loss. Reduction of optical losses in MO devices is a means to reduce their overall power consumption.

In a resent work [64], we designed and optimized using finite-difference time-domain simulations, an MPC which could enhance FR and perform high-contrast modulation simultaneously at red, green, and blue (RGB) wavelengths. This design had the structure of $(H/L)^3/(D/L)^3/(H/L)^3$ as illustrated in **Figure 5a**, where H, L, and D were chosen as $TiO_2$, $SiO_2$, and $Bi_1Y_2Fe_5O_{12}$, with optimized thicknesses of 50, 100 and 110 nm, respectively. Transmission spectrum of this MPC is shown in **Figure 5b** with three transmission peaks at 494 nm (blue), 541 nm (green) and 630 nm (red), yielding FR values of 20º, 55º, and 30º, respectively.

Fabrication of an optimized magnetophotonic crystal is challenging since it requires deposition of MO films on non-garnet substrates. This makes it highly difficult to grow single crystal MO layers. Moreover, lattice mismatch between layers can cause defects, cracks, and appearance of undesired absorptive phases. All these problems deteriorate the MO and MPC properties.
Another disadvantage of this approach is the bandwidth limitations originating from the transmitted linewidth of the magnetophotonic defect state. While higher number of cavity layers in the MPC enhances the strength of FR, as the linewidth of the transmission peak decreases, the operation bandwidth for this system also decreases. Therefore, for broader band operation, nonresonant or multiple-resonant device architectures need to be developed.

### 3.2. Magnetoplasmonics

Another major approach proposed for enhancing MO effects is magnetoplasmonics where plasmonics hybridized with MO materials enable highly localized field enhancement arising from surface plasmon resonances (SPR) [5],[6],[89],[100],[165],[166],[167],[168],[169]. **Figure 6** schematically shows a magnetoplasmonic structure and cross-sectional field profiles for such a structure, which reveal localized field enhancements in the MO film under plasmonic layer.
In this approach, localization and enhancement of the EM fields, which increases the light-matter interactions and leads to enhancement of MO effects, takes place in small subwavelength volumes near the plasmonic structures [170], and this limits the total achieved MO enhancement. Thus, enhancement happening all over a thicker film and obtaining a high total rotation requires embedding multiple layers of plasmonic structures in the film [171]. This involves challenging fabrication and significant reduction in transmission because of the optical losses from multiple layers of metallic arrays.
Plasmonic enhancement of MO effects can be particularly useful for sensing applications. Both the localized field intensity and optical activity enhancement due to plasmonic modal hybridization can help sense any minor changes in the near-field environment and identify the present materials by recognition and amplification of their fingerprint characteristics in interaction with an EM wave. In magnetooptical surface plasmon resonance (MOSPR) sensors, the enhancement of MO effect and signal-to-noise ratio enable better limit-of-detection (LOD) [172][173][174].

Raman spectroscopy is a functional method to acquire specific information about materials based on their vibrational, rotational and translational modes of bond structures and densities. This technique, however, suffers from weak Raman signal intensity which limits its sensitivity, and requires long accumulation times and large sample amount [175],[176]. Taking advantage of plasmonics in Raman spectroscopy allows for enhancement factors [177] up to $10^{14}$-$10^{15}$ and could become a solution for the mentioned drawbacks. In addition, presence of plasmonic surfaces brings the advantage of selectivity to particular analytes. Hence, surface-enhanced Raman Spectroscopy (SERS) and surface-enhanced Raman optical activity (SEROA) facilitate sensing ultralow concentrations and trace detection, especially in biological solutions, down to single-cell or single-molecule level. These techniques have gained attention in many disciplines including analytical sciences [176], chemistry and monitoring reactions [178], biomedical and pharmaceutical fields [179],[180],[181],[182], and forensic sciences [183]. Plasmonic structures serving for this purpose are normally made of noble metals (typically silver or gold) in various forms including colloidal nanoparticles [184],[185],[186],[187],[188],[189],[190], encapsulated and functionalized nanoprobes [191],[192],[193],[194],[195],[196],[197], patterned and nanostructured substrates [198],[199],[200],[201], films and roughened electrodes [202], nanoshells [203],[204], bi-metal nanoparticles (silver coated gold nanoparticles or inverse) [205],[206], controlled nanoparticle clusters [207], and immobilized metal nanoparticles on solid surfaces [208]. Localized surface plasmon resonance peak wavelength and the enhancement factor of plasmonic structures depend on composition, size, shape, proximity, and the surrounding medium of these structures [209], so the design and optimization of novel plasmonic configurations can always improve their functionality and application.

In summary, MO structures functionalized with plasmonic surfaces provide highly localized fields and significantly enhance the signal, and this makes them ideal for sensing applications. Furthermore, metallic layers could serve the dual purpose of bias contacts and magnetoplasmonic surfaces in active devices. Applications such as telecommunications could benefit from voltage control and encoding of digital ON/OFF states in the magnetoplasmonic layers. On the other hand, there are some disadvantages in using plasmonics that limit their widespread applications in practical devices. Field enhancement is only localized in small volumes around plasmonic structure. Moreover, the enhancement strongly depends on the plasma resonances of available metals, which fixes the wavelength ranges of operation and limits flexibility. Patterning plasmonic structures could be an expensive process, and these metallic structures cause high optical losses in transmission based devices.

## 4. Magnetization switching and MOSLM driving systems

In this section, we review different mechanisms reported for magnetization switching that can potentially be used as driving systems for active MOSLM devices. At the end of the section, we will give a comparison of these mechanisms and evaluate their applicability in practical MOSLMs.

### 4.1. Thermomagnetic switching

In 1958, Mayer showed the feasibility of thermomagnetic writing on the magnetic films using a heated pen [210] and an electron beam [211]. In this approach, local heating of a certain spot on a magnetic film with normal magnetization (which must be the direction of its easy axis) will cause that spot to reach the Curie temperature ($T_C$) or any suitable transformation temperature that makes the spot nonmagnetic. After cooling down below this temperature, when the spot becomes magnetized again, its magnetization would be in the opposite direction. This magnetization switching arises from the fact that thermodynamically-driven magnetic energy minimization requires flux closure through the temporarily nonmagnetic spots and, consequently, necessitates the reversal of magnetization in those spots. Thus, a thermomagnetic switching without any external bias field was demonstrated.

In the following years, Fan et al. [212],[213] used a laser (as the heating source for thermomagnetic writing) with assistance from a bias magnetic field in the opposite direction in order to reverse the magnetization of spots smaller than 3 μm in diameter, and established a magnetooptical hologram using this system. Krumme et al. [214] explained the local thermomagnetic switching by nucleation and domain wall motion. Considering strong dependence of nucleation threshold on uniaxial magnetic anisotropy ($K_u$), and the fact that in the heated region, $K_u$ is reduced and can even change sign due to a lattice misfit arising from light-induced thermal gradient, they indicated that spontaneous switching can happen with zero or small "tipping fields."

Recently, Stanciu et al. [215] experimentally demonstrated reproducible field-free magnetization reversal using a circularly polarized laser pulse. They were able to reverse magnetization of an amorphous ferrimagnetic alloy, GdFeCo, utilizing a single 40-femtosecond laser pulse with circular polarization. Two cooperating effects are involved in such a laser-induced magnetization reversal. First, ultrafast heating of the magnetic system to just below $T_C$ by absorbing part of the pulse energy, which takes it to a highly nonequilibrium state. Second, the inverse Faraday effect that causes a circularly polarized light act as a magnetic field parallel to its wavevector. The combination of these two effects allows for field-free reversal of magnetization by a laser pulse. Such a non-precessional mechanism for magnetization reversal was confirmed in subsequent studies [216]. The evolution of magnetization reversal in this mechanism is described in **Figure 7**. The images show the magnetic domains in a $Gd_{24}Fe_{66.5}Co_{9.5}$ sample with initial upward (white) or downward (black) magnetization, after excitation by 100 fs RHCP ($\sigma^+$) or LHCP ($\sigma^-$) laser pulses. In the first few hundreds of femtoseconds, pulses with both helicities take the initially magnetized material to a strong nonequilibrium state with no measurable net magnetization. The following few tens of picoseconds lead to either relaxation of the material to the initial state, or formation of a tiny domain with reversed magnetization, as seen in the last column. Helicity of the pump pulse defines the final magnetization state.

The first MOSLM [156] was composed of a garnet film and a Cu-doped CdS photoconductor sandwiched within two transparent electrodes. This so-called magnetooptic photoconductor sandwich (MOPS) was thermomagnetically switched using a HeNe laser with 1 μW power, under a 100-Oe sinusoidal magnetic field and an electric field of $1.2\times10^{4}$ $V\cdot cm^{-1}$. Thermomagnetic writing in MOPS was implemented by applying an electric pulse to the transparent electrodes and developing Ohmic heat in the illuminated area of the photoconductor, where conduction electrons are generated. Recently, Takagi et al. [217] reported an MOSLM with submicron pixels for 3D imaging in which pixels were controlled thermomagnetically with an optical addressing method. This MOSLM's magnetic pixel array that was written on an amorphous TbFe film with 10 ns pulses of a 532 nm laser is shown in **Figure 8**. In a later study [218], using an MPC structure and the light localization effect, the energy density of light required for switching was reduced by 59% compared to that of a single layer MO film.

### 4.2. Nonthermal all-optical switching

We discussed above the magnetization switching caused by the heating effect of a laser pulse that its polarization in not determinative. Using thermal effects of a laser leads to low rates of manipulating magnetization, since the repetition frequency would be limited slow cooling rates. Nonetheless, an ultrafast laser pulse is capable of manipulating the magnetization via nonthermal effects as well. Such nonthermal interactions are instantaneous and their time limitation comes only from the laser pulse widths. In contrast to thermal effects, polarization of the exciting laser plays an important role in occurrence of these effects, and they are recognized in two different types: photomagnetic and optomagnetic effects. Photomagnetic effects rely on the absorption of photons leading to an effective excitation of the magnetic system.

Optomagnetic effects are based on a coherent Raman-like optical scattering with no need for absorption of photons. In fact, optomagnetic effects are the inverse of magnetooptical effects [219],[220],[221] and can be enhanced considerably using magnetophotonic microcavities [222].

Kimel et al. [221] showed that a 200 fs circularly polarized laser pulse can act as a magnetic field pulse with amplitudes up to 5 T, and experimentally demonstrated pure optical control of spin oscillations in $DyFeO_3$ by nonthermal effects.

Hansteen et al. [223] demonstrated the feasibility of an all-optical magnetization switching happening in femtosecond time scales. They showed that both linearly and circularly polarized light can modify the magnetocrystalline magnetic anisotropy via a nonthermal photomagnetic effect and thus, establish a new equilibrium state for the magnetization. This long-lived change of the magnetocrystalline anisotropy arises from the optically induced electron transfer between ions on nonequivalent sites in the lattice, and consequent redistribution of ions in the crystal. In case of circularly polarized pulse, in addition to the described effect, a strong transient magnetic field is created along the propagation vector, through inverse Faraday effect. Although Hansteen et al. achieved only 0.6º of magnetization switching with a 100 fs pump pulse, an adequate association of optically induced magnetic anisotropy and magnetic field enables full control and reversal of magnetization.

Recently, Stupakiewicz et al.[224] accomplished an ultrafast nonthermal photomagnetic switching in Co-substituted YIG. They used a 50 fs linearly polarized laser pulse to completely steer the magnetization and showed reversible switching between two magnetic states (making possible magnetic writing and erasing) by adjusting polarization of the laser pulse. The observed initial domain structure comprised small labyrinth-like domains (black domains in **Figure 9a**) inside larger background domains (white domains in Figure 9a). $Co^{2+}$ and $Co^{3+}$ dopant ions replacing $Fe^{3+}$ in YIG result in strong magnetocrystalline and photoinduced magnetic anisotropy. Indeed, light can pump particular d-d transitions in Co ions and the resultant photoinduced magnetic anisotropy can lift the degeneracy between two metastable magnetic states in YIG:Co. Therefore, when the initial arrangement of the magnetic domains is pumped by a single laser pulse with polarization in [100] direction, large white domains ($M^{(L)+}$) simultaneously turn into large black ones ($M^{(L)-}$), and small black domains ($M^{(S)-}$) turn into small white ones ($M^{(S)+}$) as shown in **Figure 9b**. Now, a single laser pulse polarized along [010] axis that has a polarization perpendicular to the first pump pulse (**Figure 9c**).

Stupakiewicz et al. also studied the effect of pump fluence on the switched area as described in **Figure 10a**. They observed that the minimum demanded pump fluence for magnetic recording in YIG:Co is highly dependent on the wavelength of the pump pulse, as demonstrated by the data points guided with black and blue dashed lines. Examining the spectral dependence of switched area in the wavelength range of 1150-1450 nm (1.08-0.86 eV), where the electronic d-d transitions in Co ions can resonantly get excited, revealed a pronounced resonant behavior around 1305 nm (0.95 eV) as shown in the inset of Figure 10a.

Finally, the time-resolved single-shot MO images at different time lags between the pump and probe pulses ($\Delta t$), taken by Stupakiewicz et al., showed that the switched domain appears within a characteristic time $\tau$ around 20 ps and becomes stabilized after almost 60 ps, as can be seen from **Figure 10b**. The plot in this figure shows how the magnetization projection on the [001] axis (normalized to the saturation magnetization), $M_z$, changes with time. The magnetization trajectory between two states is shown schematically in the lower inset of the figure.

In contrast to optical magnetization switching in metals which basically involves thermomagnetic effects and was described in section 4.1, in transparent dielectrics, all-optical magnetic recording does not necessitate heating the medium to Curie temperature and destroying the magnetic order.

### 4.3. Current-induced Oersted field switching

In 1983, LIGHT-MOD (Litton iron garnet H-triggered magneto-optic device) [20] was introduced as an electrically-addressed reusable nonvolatile SLM with high speed and modest cost. This device comprises a bismuth-substituted iron garnet film grown on a nonmagnetic substrate and patterned to isolated pixels as can be seen in **Figure 11a**. Drivelines as an XY matrix or crossbar array were deposited and patterned using conventional optical lithography. To switch a pixel, current is passed through two adjoining row and column drivelines intersecting at the selected pixel. The magnetic field generated by only one current line is not sufficient to induce magnetization reversal in the pixels, thus an entire row or a column would not be switched. However, the combined magnetic field produced by the two of drivelines switches the magnetic state of the selected pixel. **Figure 11b** schematically illustrates the position of the drive line with respect to pixels and how switching proceeds. Like the mechanism mentioned in thermomagnetic method, switching in this method also occurs in two steps: nucleation of oppositely magnetized domain in the corner near intersection of the current-carrying drivelines, and then propagation of the domain wall to the opposite corner of the pixel and completion of switching.

Due to its electrical addressing, LIGHT-MOD was a programmable, fast and stable device and found different applications in optical processing [225]. Despite these advantages, it demanded relatively large current to generate the magnetic field required for switching the pixels [65]. Later studies tried to lessen the current requirement of this MOSLM by working on the size, shape, material, and position of the drive lines [21],[51],[65], the formation of pixels [37],[38],[51],[226], and the timing diagram for driving scheme of conductor lines [227]. These efforts resulted in reducing the drive line current from over 100 mA to below 10 mA. **Figure 12** shows the schematic of a current-driven MOSLM, which works in reflection mode and has the capability of generating a homogeneous magnetic field over the pixels owing to the design of the drive lines.

**4.4. Spin torque switching**

Another current-induced mechanism proposed for magnetization switching and potential for MOSLMs is the use of spin torques. This novel method involves no magnetic field, so it can eliminate the crosstalk problem associated with stray fields which becomes significant when the pixel pitch is small. There is no need for an active-matrix driveline system in this method, and its current demand is orders of magnitude lower than the magnetic-field-based control [52],[228]. This method provides a sub-nanosecond electrical switching for magnetic and spintronic devices [229],[230].

### *4.4.1. Spin-transfer torque (STT)*

One approach under this category is called spin-transfer torque (STT), in which the spin-polarized nature of a current passing perpendicular through a magnetic multilayer creates a spin angular momentum transfer between the magnetic sublayers. Devices consisted of altering magnetic and nonmagnetic metal layers of a few nanometers thicknesses show giant magnetoresistance (GMR) [231], where current flow is strongly affected by alignment of magnetic moments in the layers. The inverse effect can also be expected where electrons scattering in the device affect the magnetic moments in the layers by applying torques on magnetic moments and transferring angular momentum between layers. This so-called spin-transfer torque (STT) is reported for different material systems [232],[233],[234],[235],[236] and its compatibility with MO devices is demonstrated [52],[74],[237],[238]. **Figure 13** shows the cross-sectional schematic of a single pixel in a spin-transfer-switching MOSLM. The pixel structure consists of two magnetic layers separated by a spacer layer, which can be a non-magnetic metal (where the stack is called spin valve) or insulator (where the stack is called magnetic tunnel junction) [239], a bottom electrode, and a transparent electrode on top. In one of the magnetic layers, named as free layer, magnetization can change by a small magnetic field, while in the other one, called pinned layer, a large magnetic field is needed for switching the orientation of the magnetization.

One of the magnetic layers designated as free layer changes magnetization with a small magnetic field, while the other magnetic layer, called pinned layer, requires a large magnetic field to switch the magnetization orientation. In order to use such a structure for MO modulation of light, the free layer needs to have MO properties. When a current applied perpendicular to the films plane, passes through the pinned layer, it gains spin polarization and when this spin-polarized current is directed to the free layer, its angular momentum can be transferred to the free layer, changing its magnetization direction.

Typical current densities required for magnetization switching in this method are on the order of $10^7$ $A.cm^{-2}$ [239].

*4.4.2. Spin-orbit torque (SOT)*

The other approach that exploits spin torques for magnetization switching is called spin-orbit torque (SOT). This mechanism involves spin Hall [228], Rashba [240],[241] or Dresselhaus [241] effects and benefits from the coupling between spin and orbital motion of electrons to create a non-equilibrium spin accumulation, which subsequently applies a torque on magnetization by means of spin transfer. As a fundamental property of spin-orbit (SO) coupling, motion of electrons under an electric field is accompanied with a magnetic field, called SO field. Even if no external magnetic field is present, SO field couples to the magnetic moment of the moving electrons [242].

According to the spin Hall effect (SHE) [243], when a charge current is flowing, SO coupling causes the electrons with spin up to deflect in one direction perpendicular to the current path and the electrons with spin down in the opposite direction. As a result, an unpolarized charge current converts to a pure spin current in transverse direction. Since the number of spin up and spin down electrons are equal in an unpolarized current, a net charge flow perpendicular to the applied current would not form. This concept is schematically shown in **Figure 14**. The spin current can be utilized to modify the direction of magnetization in an adjacent layer, by applying a spin-transfer torque as explained in the previous section.

According to Rashba effect, when a ferromagnetic film is sandwiched between two dissimilar materials, the electric potential is highly asymmetric in the direction perpendicular to the films plane, which results in a structural inversion asymmetry (SIA) in this direction. Electrons moving in such a structure experience a net electric field, E, that transforms to an effective magnetic field, $H_R$, due to SO interaction:

$$H_R = \alpha_R (\hat{z} \times \langle k \rangle) \quad (4)$$

where $\alpha_R$ is a material parameter related to the strength of the SO coupling, $\hat{z}$ is the unit vector parallel to E and $\langle k \rangle$ is the average electron wavevector. When no current is applied, populations of the k and −k states are equal and $\langle k \rangle = 0$, however, when a charge current is applied, distribution the electrons in k-space becomes asymmetric. This produces a net effective field and creates a non-equilibrium spin accumulation perpendicular to the current flow, which can subsequently apply a torque on magnetic moment of the material and induce magnetization reversal [244],[245].

In a similar effect, if the electric field E resulting in an effective magnetic field and thus, a spin-orbit torque, is due to a bulk inversion asymmetry (BIA), this mechanism is called Dresselhaus effect [241] . The Dresselhaus effect is observed in crystals with zinc blende structure that lacks an inversion center.

Typical heterostructures exhibiting SOTs consist of a ferromagnetic film sandwiched between a heavy metal with strong SO coupling, and an insulator [246],[247]. **Figure 15** shows the Rashba field produced by a charge current in such a structure. This approach has been extensively studied and improved for spintronic devices [248],[249],[250],[251],[252],[253],[254],[255],[256],[257],[258],[259].

### 4.5. Multiferroic switching

In spite of the above-mentioned improvements in current-driving of MOSLMs, the required current densities (typically ranging from $10^5$ to $10^7$ $A.cm^{-2}$) still exceed the critical current values for commercially acceptable energy consumption and reliable device operation. Because of the current passing through the conductors with small cross section, Joule heating occurs over small pixels and causes switching errors due to the temperature drift. Thermal drift due to Joule heating is inevitable in the current-based methods. This situation has motivated the search for lower power switching schemes.

Early experimental demonstrations have shown that by eliminating resistive losses, multiferroics reduce the energy dissipation per unit area per switch to 1-500 $\mu J.cm^{-2}$, which is a noteworthy advancement in comparison to that of current-driven switching (1-10 $mJ.cm^{-2}$) [260]. Multiferroic materials are compounds in which several ferroic orders coexist. These orders can be ferromagnetic and ferroelectric for instance, where the coupling is called magnetoelectric effect. For practical device purposes, electrical control of magnetization is desired and this fact implies the importance of magnetoelectric effect among multiferroics. Multiferroic magnetoelectric structures can be divided into two categories: single-phase multiferroic materials in which magnetoelectric coupling enables tailoring the magnetic properties by application of an electric field, and composite structures which combine a ferroelectric (or piezoelectric) material with a ferromagnet (or magnetostrictive component). In practice, the magnetoelectric coupling in composite systems happens through one of the following physical mechanisms [70]:

- Strain coupling: modification of the magnetic properties through magnetostriction, which is controlled through voltage-driven strain changes [261],[262].
- Impact of the polarization direction of the ferroelectric component on the electronic structure of the ferromagnetic one in their interface [263].
- Harnessing the exchange interaction between a ferromagnet and a (magnetic) ferroelectric material [264].

**Figure 16** schematically shows a multiferroic voltage-driven MOSLM that uses a piezoelectric material, PZT [$Pb(Zr_{0.52}Ti_{0.48})O_3$], for magnetoelectric coupling. By applying a voltage through the crossbar metallic contacts, over a pixel with out-of-plane magnetization, a stress is created by the electrostrictive PZT layer on the selected pixel. This stress acts as an effective field which helps the magnetized domains to realign to the film plane due to magnetostriction effect. In this stage, a small bias field in the direction opposite to the initial magnetization causes the selected pixel to easily switch and retain that state even after turning off the voltage.

Magnetoelectric or voltage-driven magnetization reversal minimizes resistive losses and in principle, only needs a charge supply sufficient for a charge/discharge of a capacitor. This reduces the energy dissipation by two orders of magnitude in comparison to the current-driven methods [265] and makes it a superior candidate for switching pixels in MOSLMs. This mechanism has been experimentally reported to work through modification of magnetic anisotropy [101],[262],[263],[266],[267],[268],[269], changing magnetic state [270],[271],[272], or coupling of ferroelectric and magnetic domains [273],[274].

In summary, various magnetization switching methods reviewed in this section have their advantages and disadvantages for application in driving system of MOSLMs. Among all these methods, voltage driving which is a multiferroic switching mechanism, seems more viable for industrializing MOSLMs, as it provides an electrical control of device without substantial Ohmic power loss and thermal drift. Even this method has its drawbacks as it requires materials and structures which can provide both high MO quality and good magnetoelectric coupling. This normally happens through strain coupling, which involves piezoelectric materials and has its own complications. All these challenges that have stalled commercialization of MOSLMs by now, necessitate finding novel materials and driving methods. **Figure 17** shows different magnetic properties that can be controlled by electric field. These properties can influence magnetization direction and, therefore, can be considered as potential mechanisms for low-power magnetization switching.

## 5. Open challenges and research opportunities in MOSLMs

In spite of decades of research regarding magnetooptical spatial modulation of light, there are still challenges and questions in the field that need to be addressed to achieve functional high-performance MOSLMs.

Developed SLM products and MOSLM lab prototypes were compared in Tables 1 and 2, respectively. Liquid crystal-based products typically consume more than 10 W for 60 Hz frame rate and resolutions greater than 1024×1024 (1K×1K) pixels, which are necessary for most applications. These devices require voltage control of the orientation of liquid crystals with micron-scale thicknesses. Their thicknesses cannot be reduced due to the number of molecules needed for $\pi$ phase shift during modulation. As a result, the power consumption of LC-based devices cannot be reduced significantly without a major change in switching mechanism or materials. On the other hand, digital micromirror devices (DMD) provide an order of magnitude or more decrease in power consumption for comparable specifications due to the microelectromechanical actuation in DMD. The main disadvantages of DMD systems are that they are mainly binary (not grayscale) and have significantly more complex fabrication processes than LC-based devices. Mechanically moving parts in DMD may also raise reliability issues over their operation lifetimes. For MOSLMs, there are no mechanically moving parts and early MOSLM demonstrations indicate frame rates exceeding 1 kHz, which is advantageous compared to other SLM products. In MOSLM demonstrations, binary-only modulation was achieved by voltage and/or external magnetic field-based magnetization reversal. Power requirements for switching pixels are mostly below 1 W, however, device driving electronics and optics are not included in these figures. The current densities in MOSLM still need to be reduced and external magnetic field needs to be eliminated for avoiding pixel crosstalk. Spin torque switching methods have current demands ($10^5$ to $10^7$ $A.cm^{-2}$) orders of magnitude lower than the magnetic-field-based control, without a need for external magnetic field. Multiferroics reduce the energy dissipation per unit area per switch to 1-500 $\mu J.cm^{-2}$ by eliminating resistive losses, which is a noteworthy improvement in comparison to that of current-driven switching (1-10 $mJ.cm^{-2}$). Advancement in terms of pixel count and pitch sizes is necessary for commercially viable device applications. Micro/nanofabrication process developments in the last decade could help reduce the pixel sizes and the required modulation voltages, and help increase the pixel count to projector or display standards (1K×1K or above). With the development of high figure-of-merit MO materials for both visible and near infrared, the wavelengths and operation bandwidth of the MOSLMs could be extended significantly. Implementing the major materials and switching breakthroughs presented in the previous sections could help reduce the power consumption and improve the mentioned features in MOSLMs.

An ideal SLM needs to be capable of a complete amplitude modulation between zero and input light intensity which necessitates 90° of polarization rotation in MOSLMs in order to achieve a high-contrast and power-efficient device. The energy required for switching the pixels in MOSLM has to be acceptable such that competing industrialization requirements could be met. Any unnecessary power consumption like optical losses and Joule heating from pixel drive electronics should be minimized in an ideal device. Minimizing these losses not only could lead to a low-power operation but also could prevent the problems associated with thermal drift. Integration of a 2D array of pixels and addressing electronics to the device must be devised in a way that provides the necessary spatial resolution for a specific application while avoiding crosstalk between pixels. For holographic imaging and 3D displays, pixel pitches below 1 µm is preferred for compact devices. Although binary modulation meets the needs of simple displays used in calculators, e-readers, electronic labels, etc., more complicated applications including imaging and holography, require analog or multilevel modulation with sufficient modulation depth (8 bit or more). These applications are normally not satisfied with only amplitude modulation but also necessitate full complex modulation.

In order to fulfill all the above-mentioned requirements and develop the MOSLM as an advanced product for emerging applications, research in the following direction is suggested:

(i) Direct investigation of the microscopic origin of MO effects and complete understanding of the physics behind these phenomena should be thoroughly accomplished. These studies would allow for the discovery and synthesis of new MO materials in addition to novel approaches for enhancement of MO effects and might help proceed towards achieving 90º rotation and efficient modulation for a broad range of wavelength.

(ii) Integration of fast and low-power driving optics and electronics to MOSLMs is another challenge that needs to be resolved. Among the present switching systems, voltage-driving scheme is more promising because of not involving significant Joule heating and power loss, in addition to the simplicity of configuration. However, exploration of multiferroic materials and structures with strong magnetoelectric coupling becomes an open research area for voltage driving.

(iii) Forming pixels smaller than 1 µm is not a serious obstacle on the way of developing advanced MOSLMs using the advanced micro/nano fabrication technologies. Depending on the driving scheme, MO material islands may not be needed; however, scaling of drive electronics to small sizes is much more challenging.

(iv) In terms of modulation type, binary modulation can be performed in principle and has been experimentally demonstrated in MOSLMs [21],[37],[38],[65],[156],[227],[226] but analog or multilevel modulation remains a challenge due to the hysteresis behavior in MO materials. We suggest implementing analog modulation in MOSLMs using minor hysteresis loops or return paths to different remanent state as shown in **Figure 18**. In this approach, different levels of MO signal intensities are achievable by magnetizing the pixels up to different points using the minor loops, instead of full magnetic saturation. Another way is saturating all the pixels along hard magnetic axis and then, turning the driving signal (e.g. voltage) off and letting the pixels relax towards easy axis over time and stabilize the B-field using pulse width modulation schemes. Since the MO effects are proportional to magnetization, pixels with different magnetization levels lead to different rotation angles and as a result, different pixel intensities will be obtained.

(v) Phase modulation or full complex modulation using MO materials have not been proven yet. The fact that left- and right-handed circularly polarized (LHCP and RHCP) light experience different refractive indices in a magnetized MO material, can provide a platform to investigate possibility of phase control and full complex modulation using MOSLMs.

**6. Conclusion**

In this review, we described the progress on magnetooptical spatial light modulators (MOSLMs) from materials point of view to device architectures and driving systems. MOSLMs hold promise to simultaneously provide high modulation speed and fine spatial resolution, and this makes them superior over their counterparts including liquid crystal (LC) based SLMs and digital micromirror devices (DMDs). These properties in addition to nonvolatility, solid-state and no moving components make MOSLMs ideal for a variety of applications such as holography, 3D displays, ultra-broadband optical telecommunication, beam steering devices, etc., and yet, there are challenges that have hindered complete development and commercialization of these devices. In this paper, we also tried to cover all the challenges associated with MOSLMs and investigate the possible solutions in order to help guide the research in the field towards practical and functional MO devices.

MO effects have been studied in different materials among which dielectrics and especially garnets provide the most applicability for MO devices. However, these materials mostly have MO figures of merit smaller than 1 $°\cdot dB^{-1}$ which is not sufficient for realizing practical miniaturized devices. Methods such as doping garnets with Bi, Ce, and Nd, appropriate heat treatment during or after synthesis of MO films, and making composite materials are suggested for improving MO properties. Recently, giant MO effects discovered in 2D materials such as graphene and silicene have opened new prospects for the field. Rotation angles as large as 13° are demonstrated for a monolayer, in this group of materials.

In terms of device engineering, back reflection from a mirror behind MO film, magnetophotonic crystals (MPCs), and magnetoplasmonic structures have been suggested to increase polarization rotation and MO figure of merit. While using a back reflector became part of the MOSLM designs, the challenges such as complexity of fabrication, high optical losses, and narrow bandwidth held back applications of MPCs and magnetoplasmonics in practical MO devices. Ongoing research in the field is expected to alleviate these difficulties.

The demonstrated and suggested driving systems for MOSLMs include switching with thermomagnetic, nonthermal optical, current-induced Oersted field, spin-torque, and multiferroic effects. Comparing advantages and disadvantages of these methods, voltage driving which is a multiferroic switching mechanism, shows promise for MOSLMs, because it provides an electrical control of device without substantial Ohmic power loss and thermal drift. This method also has obstacles on the way of its development as it requires materials and structures which can provide both high MO quality and strong magnetoelectric coupling. These requirements pose significant material compatibility challenges.

In spite of the progress on different aspects to develop MOSLMs as high-performance device, no systematic investigation is reported on analog or multilevel modulation, which is a prerequisite for many applications. We suggested using minor hysteresis loops or pixel-wise control of time-dependent magnetization decay for this purpose.

With the development of MOSLMs as a functional device, new ultrafast, nonvolatile and high resolution capabilities in holography, heads-up displays, virtual and augmented reality devices, solid-state light detection and ranging (LIDAR), optical communications, data storage and other emerging applications can be achieved.

**Acknowledgements**
This study was financially supported by the European Research Council Advanced Grant (ERC-AdG) Wear3D Project No: 340200 and BAGEP 2017 Award, TUBITAK Grant No: 117F416. The authors gratefully acknowledge useful discussions with Prof. Mitsuteru Inoue and Dr. Taichi Goto.

**Conflict of Interest**
The authors declare no conflict of interest.

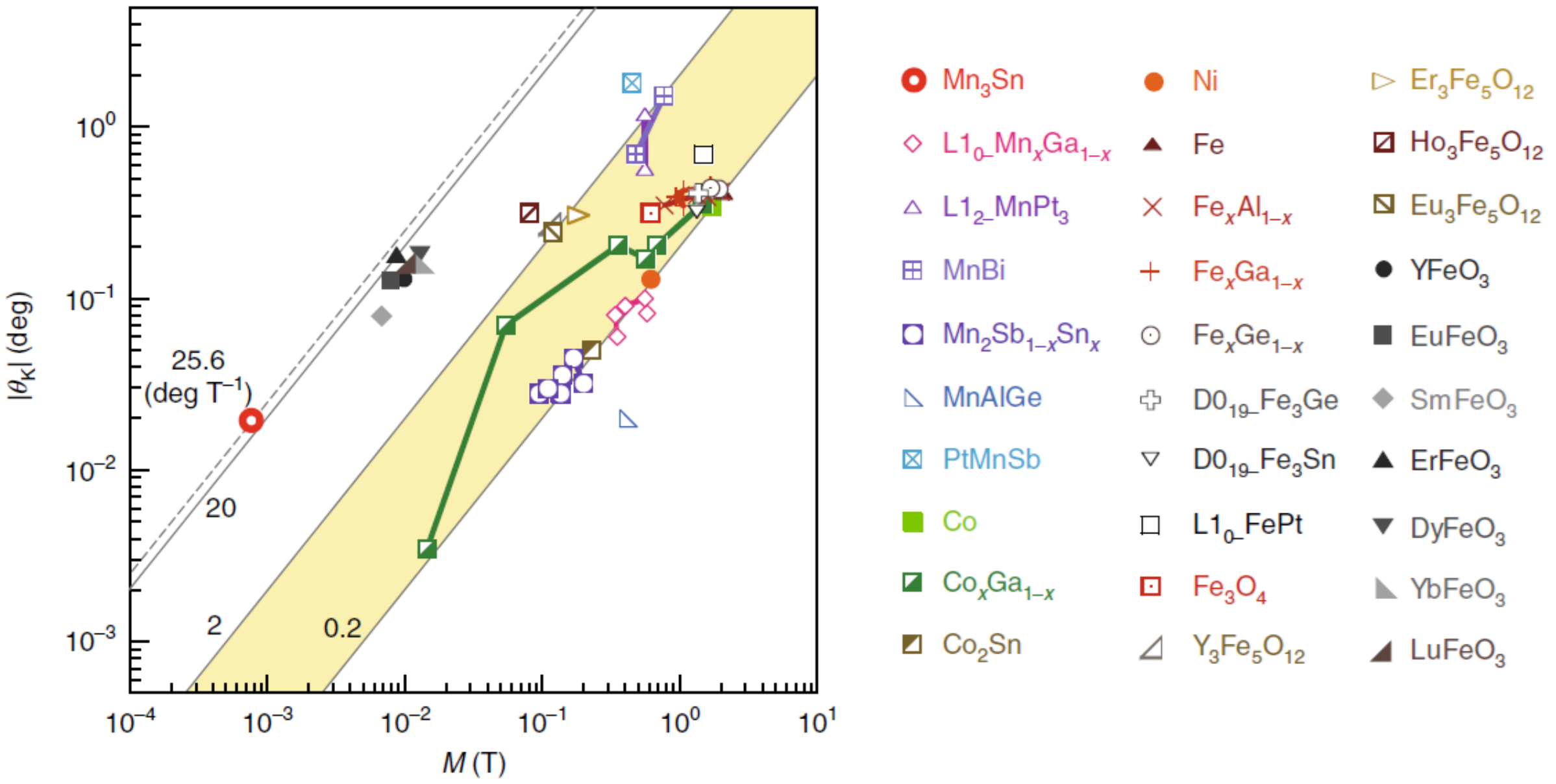

**Figure 1.** Polar MO Kerr rotation angle $|\theta_K|$ measured as a function of magnetization for various ferro-, ferri-, and antiferromagnetic materials at room temperature. For most of the ferro/ferrimagnets, $|\theta_K|$ is proportional to M, $|\theta_K| = K_s M$. $Mn_3Sn$, an antiferromagnetic metal, has a large MOKE with $K_s$ = 25.6 deg.$T^{-1}$ while for antiferromagnetic insulators, $K_s$ has a value in the range of 10-20 deg.$T^{-1}$. Reproduced with permission. [87] 2018, Springer Nature.

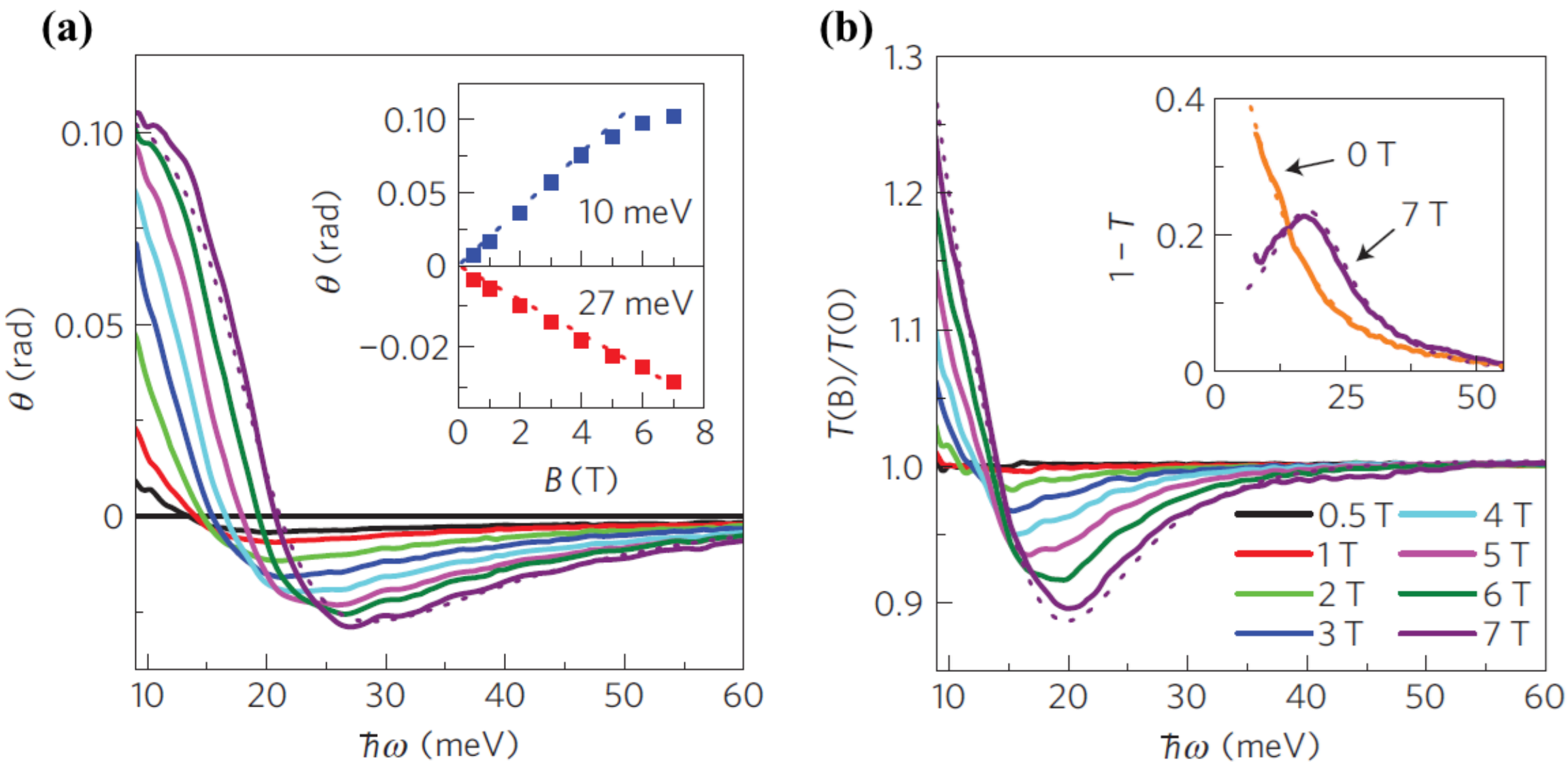

**Figure 2.** Faraday rotation and magnetooptical transmission spectra of single layer graphene. (a) Faraday angle θ measured at 5 K under different magnetic fields. The inset shows the magnetic field dependence of θ at photon energies ħω = 10 and 27 meV. (b) The zero-field-normalized transmission spectra under the same magnetic fields. The inset shows the absorption (1−T) at 0 and 7 T. Reproduced with permission. [109] 2011, Springer Nature.

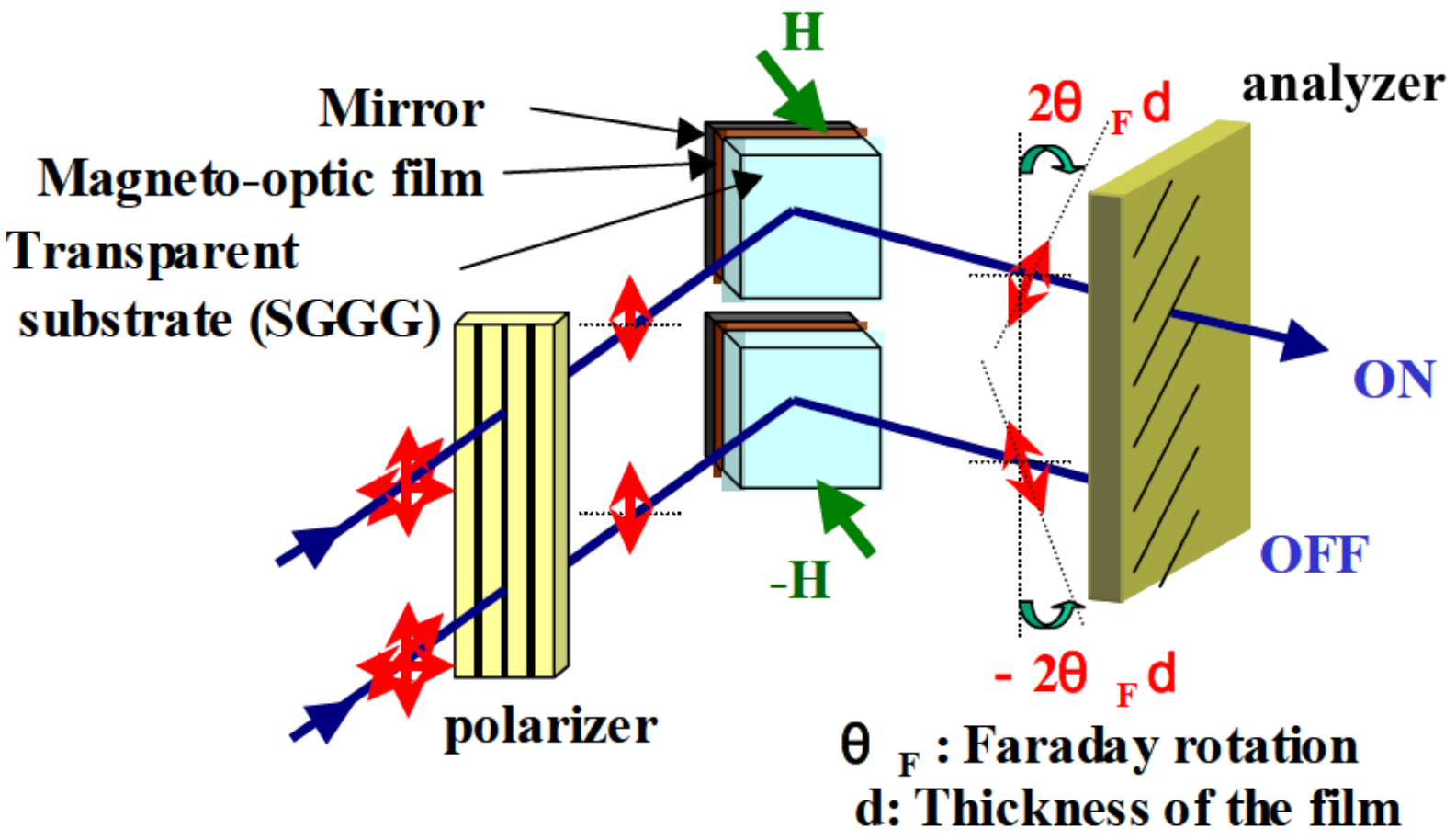

**Figure 3.** Working principle of an MOSLM which uses Faraday effect with a back-reflector to double the rotation angle. The linearly polarized light coming out of a polarizer enters from the transparent substrate and traverses the magnetooptical film, reflects back and exits from the substrate. For a film with specific Faraday rotation $\theta_F$, thickness d; the rotation angle in transmission mode is $\theta_F \cdot d$. Using a reflector causes an overall rotation angle of $2\theta_F \cdot d$ due to nonreciprocal characteristic of the MO effects. Direction of the rotation depends on the direction of applied magnetic field H. Reproduced with permission. [51] 2006, SPIE.

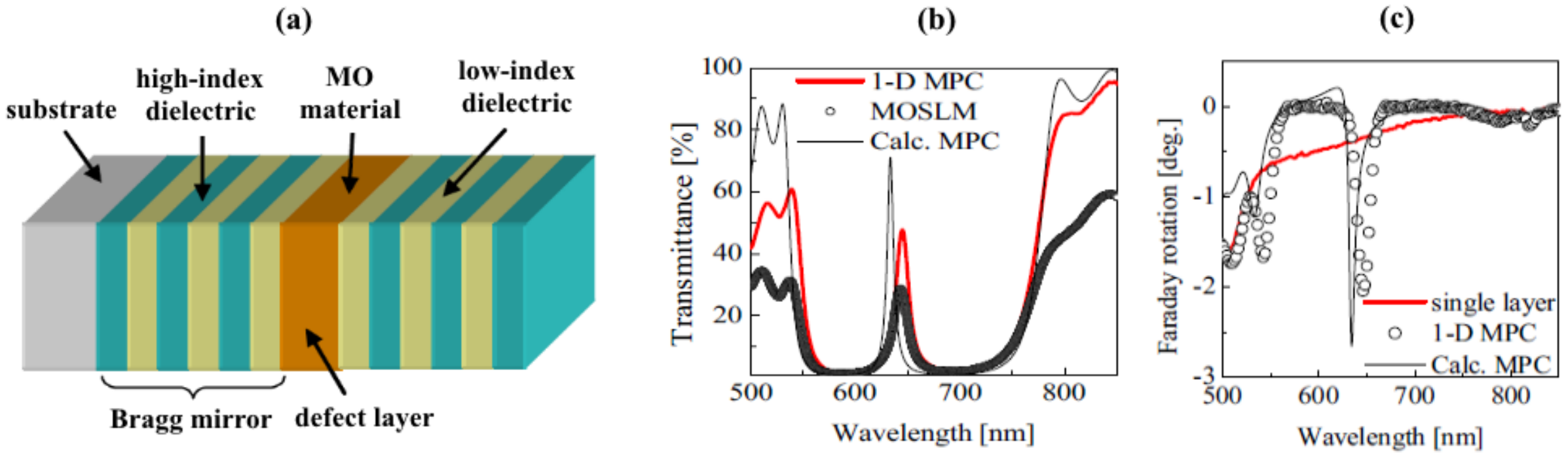

**Figure 4.** Photonic crystal-based enhancement of MO effects. (a) Schematic of a single-defect magnetophotonic crystal (MPC) with a structure in the form of substrate/$(H/L)^3/D/(L/H)^3$ where H, L and D strand for high-index dielectric, low-index dielectric and defect layer, respectively. (b) Transmission and (c) Faraday rotation spectra for a 1-defect MPC (calculated and experimental data are shown by black and red lines, respectively) and an MOSLM comprised of such MPC (data points shown by shallow dots). Reproduced with permission. [22] 2007, John Wiley and Sons.

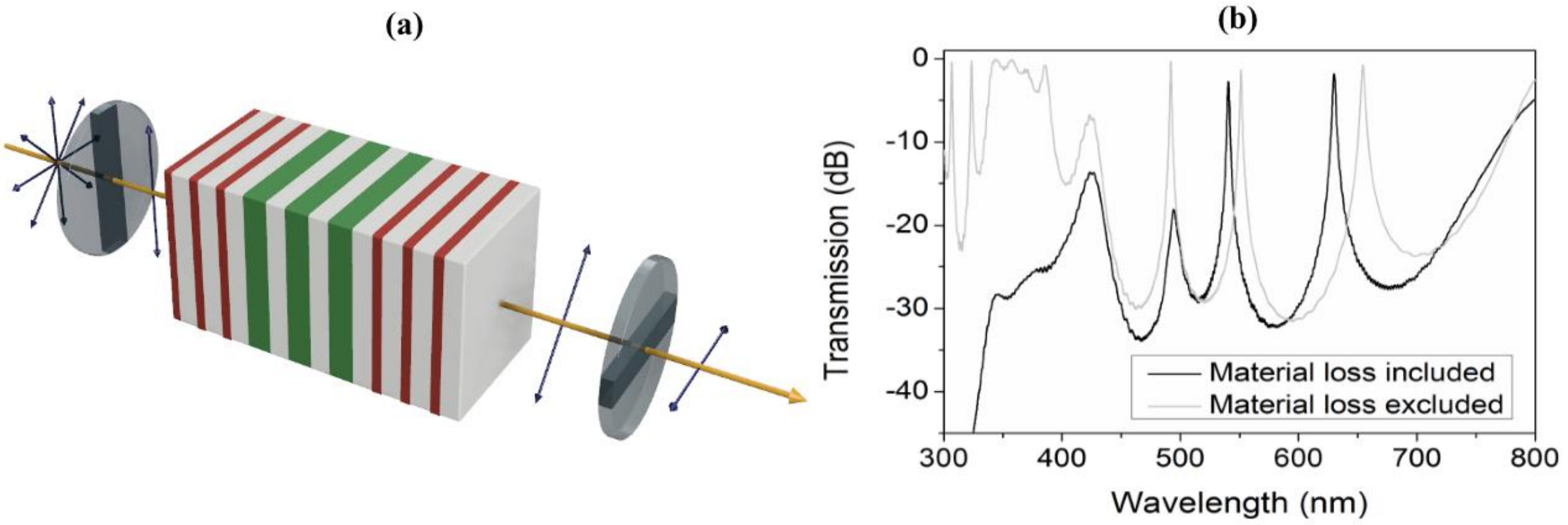

**Figure 5.** RGB magnetophotonic crystal. (a) Schematic of a three-defect MPC with $(H/L)^3/(D/L)^3/(H/L)^3$ structure designed for enhancing Faraday rotation and high-contrast modulation at three fundamental wavelengths of red, green and blue (RGB). (b) Transmission spectrum of the three-defect MPC showing transmission peaks at $\lambda$ = 494 nm (blue), 541 nm (green), and 630 nm (red). Black (gray) line shows the case where the MO material is lossy (lossless). Reproduced under the terms of the Attribution 4.0 International (CC BY 4.0) license. [64] 2019, Springer Nature.

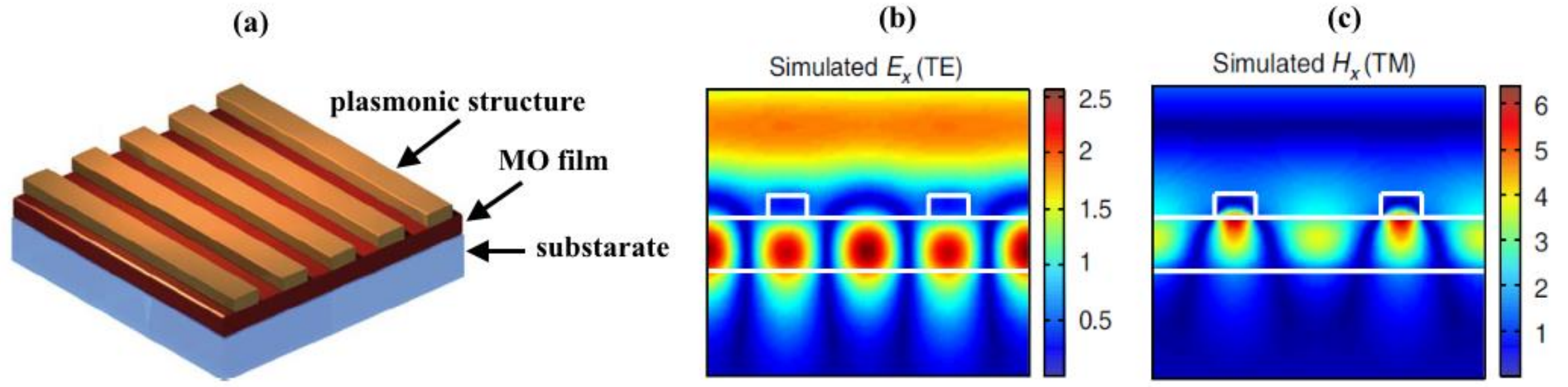

**Figure 6.** Plasmonic enhancement of MO effects. (a) Schematic illustration of a magnetoplasmonic structure, (b) electric field profile for a TE-polarized incident light, and (c) magnetic field profile for a TM-polarized incident light. Field amplitudes were calculated by numerical methods and normalized with respect to that of incident light. Reproduced with permission. [167] 2013, Springer Nature.

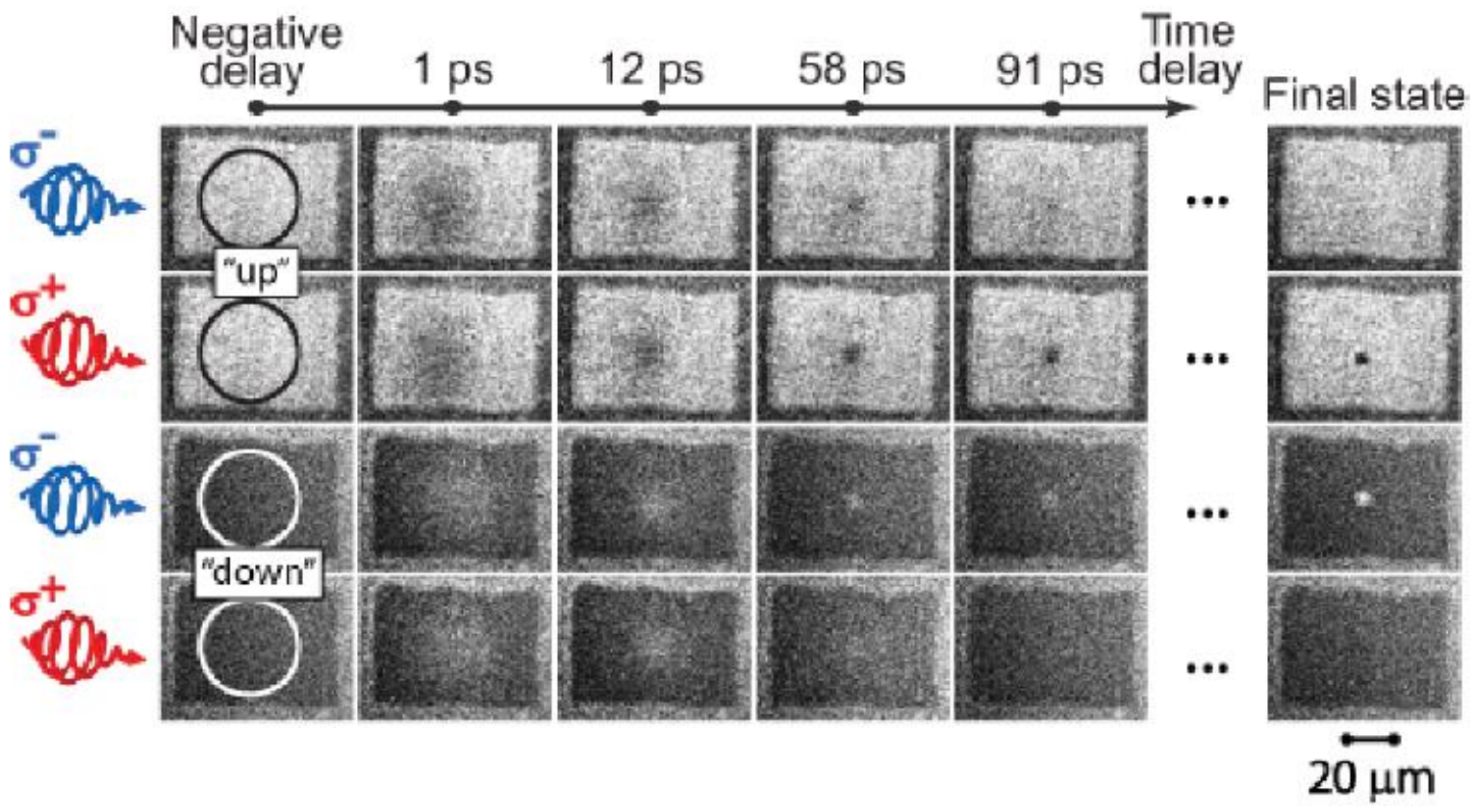

**Figure 7.** Thermomagnetic switching via a nonequilibrium state. The images show the magnetic domains in a $Gd_{24}Fe_{66.5}Co_{9.5}$ sample with initial upward (white) or downward (black) magnetization, after excitation by 100 fs right- ($\sigma^+$) or left-handed ($\sigma^-$) circularly polarized laser pulses. The circles show areas actually affected by pump pulses. In the first few hundreds of femtoseconds, pulses with both helicities bring the originally magnetized medium to a strong nonequilibrium state with no measurable net magnetization, seen as the gray area in the second column. In the following few tens of picoseconds, either the medium relaxes back to the initial state or a small domain with a reversed magnetization is formed as seen in the last column. Reproduced with permission. [216] 2009, The American Physical Society.

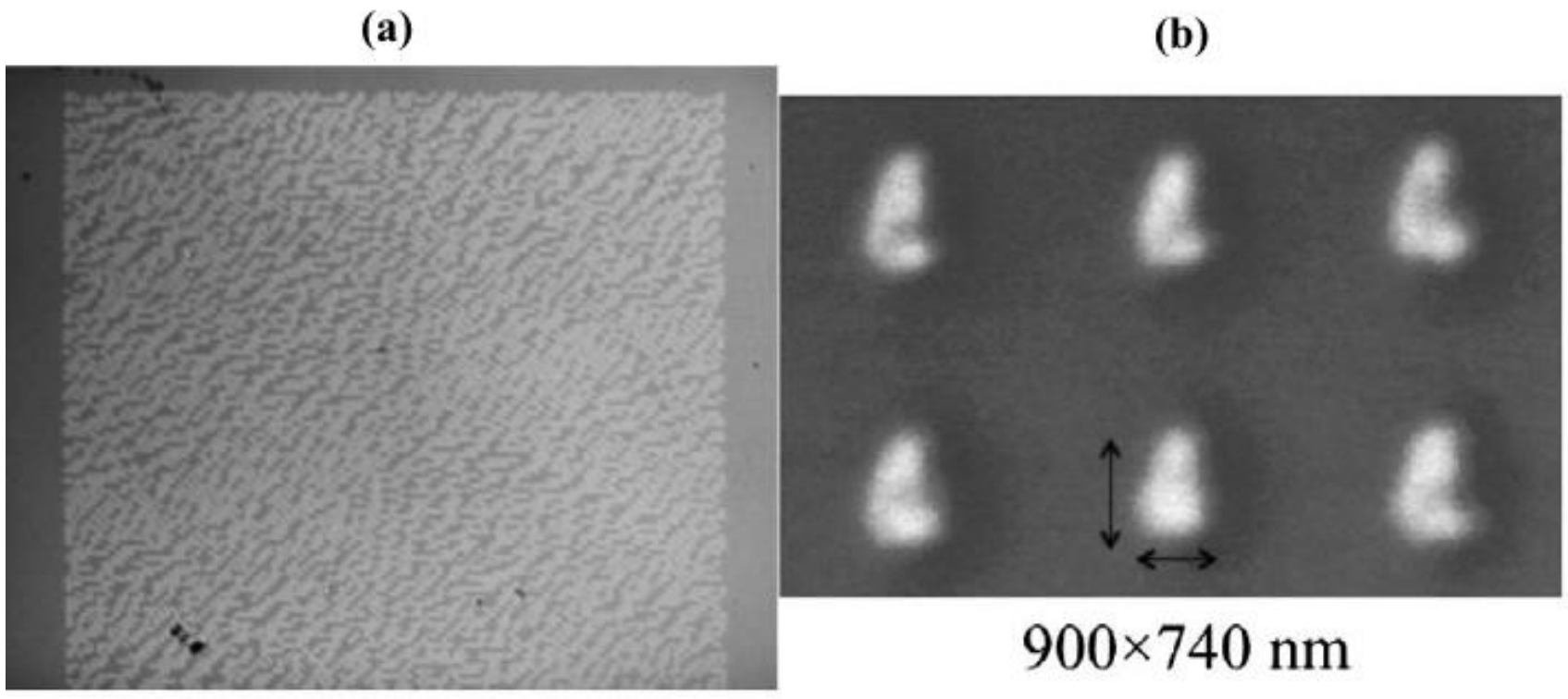

**Figure 8.** Submicron MOSLM pixel array written on an amorphous TbFe film with 10 ns pulses of a 532 nm laser. (a) Polarization microscope image of 256×256 pixels with 1 μm pitch and (b) magnetic force microscopy image of 3×2 pixels with 2.5 μm pitch. Reproduced with permission. [217] 2014, Optical Society of America.

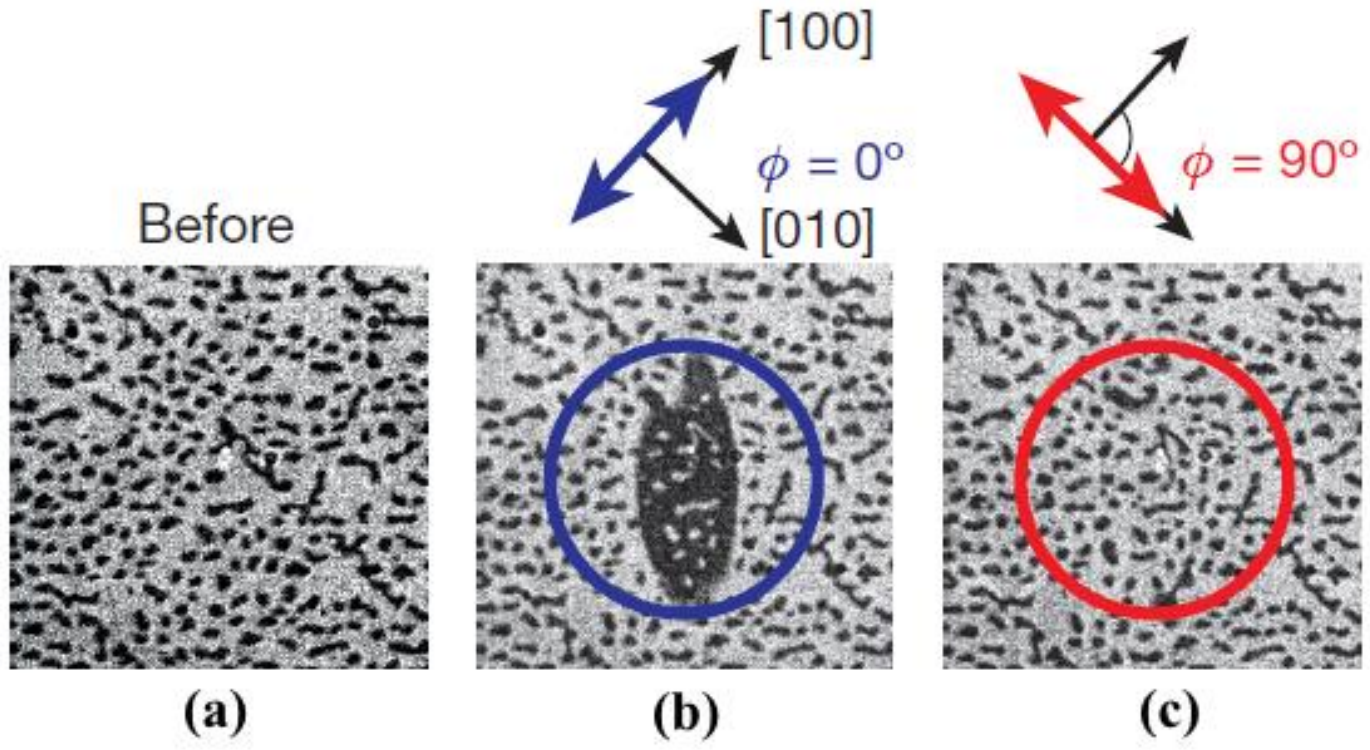

**Figure 9.** Reversible magnetization switching (magnetic writing and erasing) by nonthermal effects of a laser pulse. The pump beam with the wavelength of 1300 nm was focused to a spot 130 μm in diameter and with maximum fluence of 150 mJ·cm$^{-2}$. $\phi$ is the angle between the pump pulse polarization and the [100] axis. The images are taken by a femtosecond magnetooptical imaging technique and are 200 μm × 200 μm in size (a) Initial domain structure before laser excitation, which was prepared by applying an external magnetic field $\mu_0 H$ = 80 mT along the [1$\bar{1}$0] axis for a few seconds. (b) Domain structure after excitation by a single laser pulse polarized along the [100] axis, and (c) subsequent excitation with a similar laser pulse polarized along the [010] axis. Reproduced with permission. [224] 2017, Springer Nature.

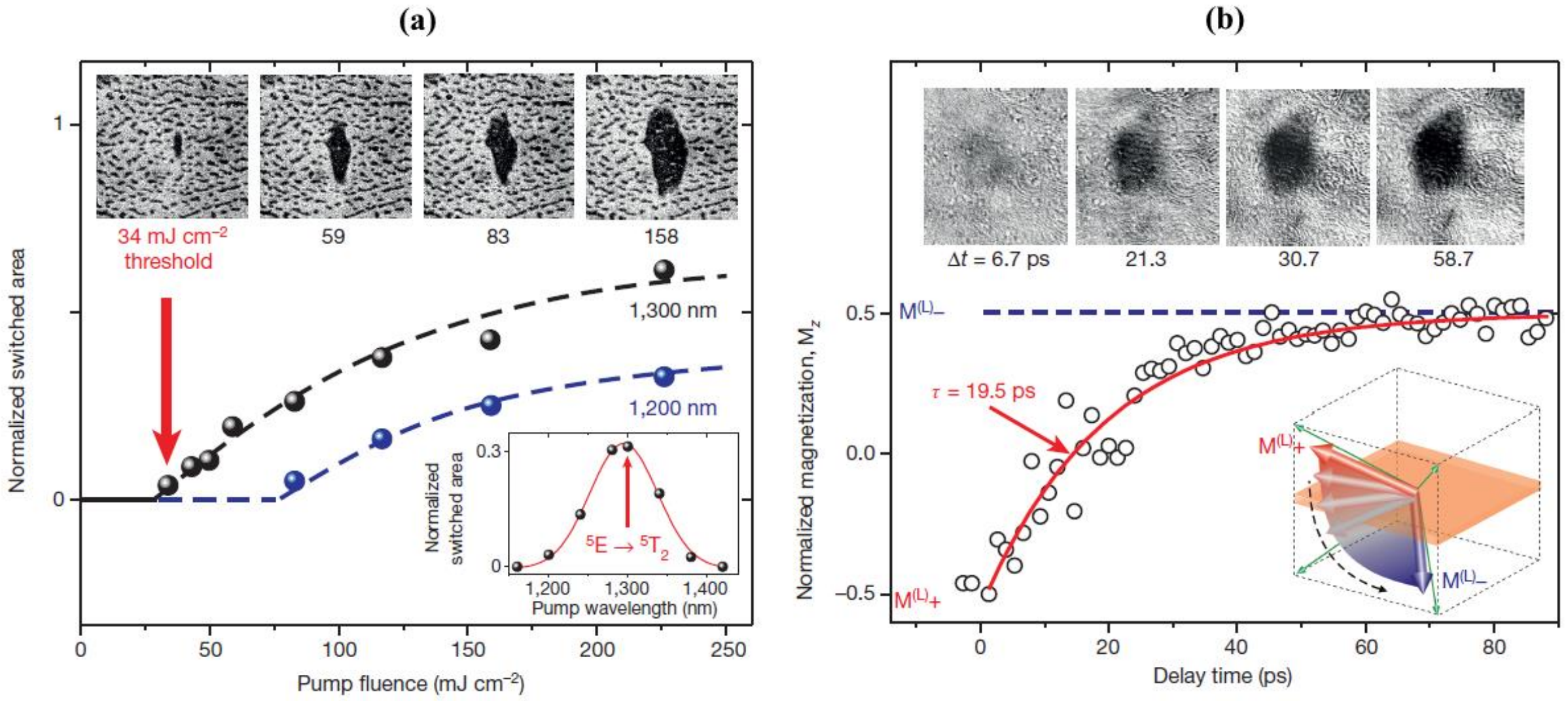

**Figure 10**. Pump fluence and time dependence of magnetization evolution in nonthermal all-optical switching. (a) The normalized switched area, calculated as the ratio of the recorded domain area (the black large domain on the images) to the area of pump laser spot $\pi r^2$ (r: the pump spot radius) plotted as a function of the pump fluence. The plots correspond to the cases when the central wavelengths of the pump are around 1200 nm (blue dots) and 1300 nm (black dots and images). The minimum pump fluence required for magnetic recording in YIG:Co is very sensitive to the wavelength of the pump pulse. The inset shows the spectral dependence of the normalized switched area for a pump fluence of 83 mJ·cm$^{-2}$. (b) Time-resolved switching process in YIG:Co observed by femtosecond single-shot magnetooptical imaging. Images (240 μm × 260 μm) are taken at different time delays between pump and probe pulses, and shown after subtraction of the reference image taken before excitation. The switched domain emerges within a characteristic time τ around 20 ps and stabilizes after about 60 ps. Time dependence of the magnetization projection on the [001] axis ($M_z$) normalized with respect to the saturation

magnetization is plotted. The data points were calculated as the ratio of the magnetooptical signal (the average image contrast) in the switched area to the magnetooptical signal in the case when the magnetization is aligned along the [001] axis. The red solid line was fitted using the exponential increase $(1 - \exp[-\Delta t/\tau])$ with the characteristic time $\tau = 19.5 \pm 1.6$ ps. The lower inset shows the schematics of the magnetization trajectory during the switching. The magnetization is switched between $M^{(L)+}$ and $M^{(L)-}$ states with the help of the laser pulse polarized along the [100] axis (pump fluence: 150 mJ·cm$^{-2}$, pump central wavelength: 1250 nm). Reproduced with permission. [224] 2017, Springer Nature.

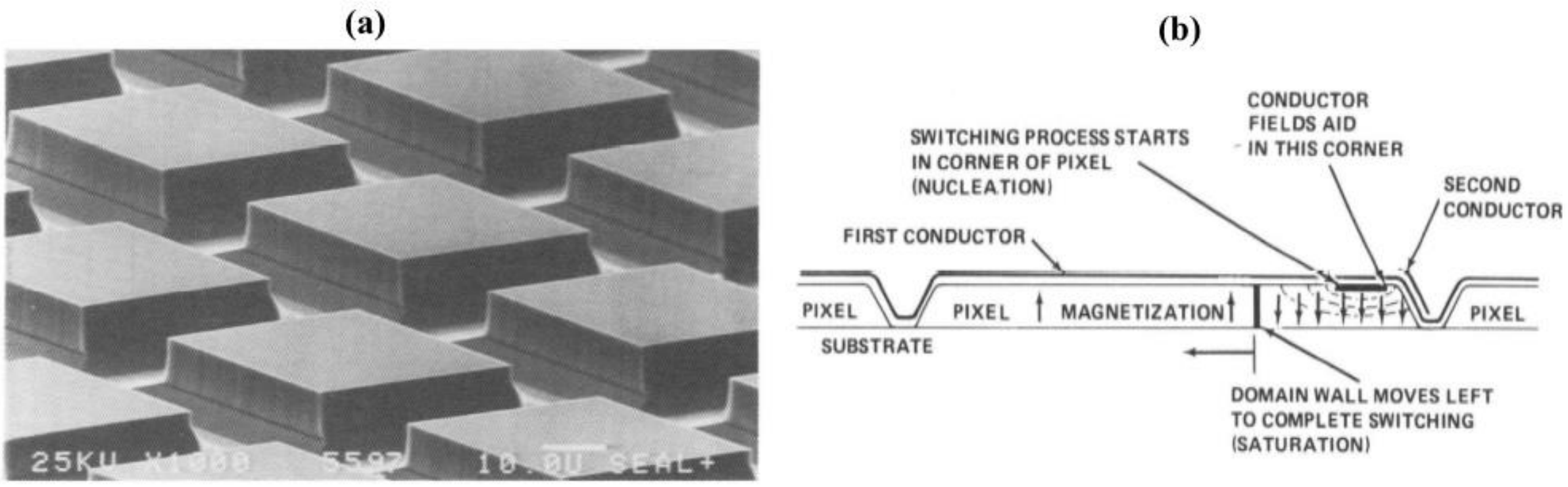

**Figure 11.** Litton electrically-addressed reusable nonvolatile SLM. (a) SEM image of the MO film patterned into pixels, (b) schematic position of conductor lines on a pixel and switching process. To switch a pixel, current is passed through the two adjoining row and column conductors intersecting at the selected pixel. First, an oppositely magnetized domain nucleates in the corner near the selected conductor intersection. Then, the domain wall propagation to the opposite corner of the pixel and completes switching. Reproduced with permission. [20] 1983, Society of Photo-Optical Instrumentation Engineers.

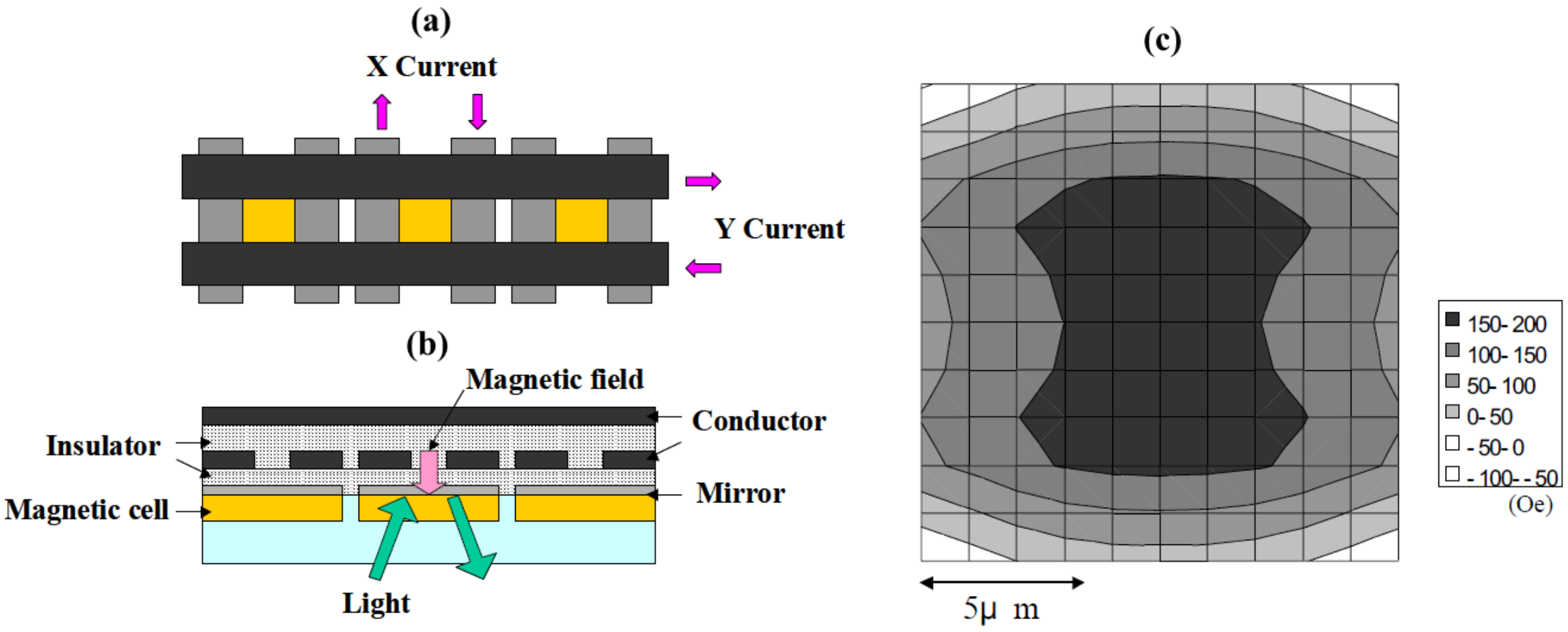

**Figure 12.** A current-driven MOSLM, working in reflection-mode and capable of generating a rather homogeneous magnetic field over the pixels owing to the design of the drive lines. Schematic (a) top and (b) side views. The conductors resemble a small coil on top of each pixel. (c) Field distribution over a pixel. Reproduced with permission. [51] 2006, SPIE.

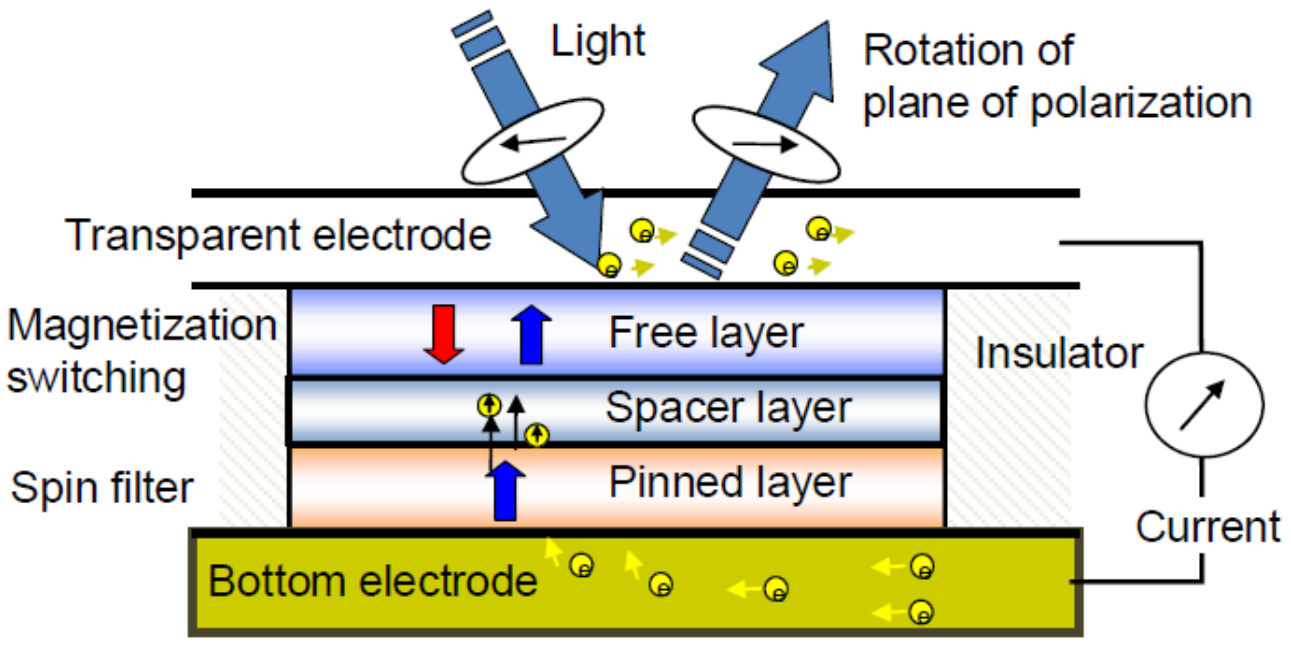

**Figure 13.** Schematic cross-section of a single pixel in a spin-transfer-switching MOSLM. The pixel structure comprises two magnetic layers separated by a spacer layer, a bottom electrode, and a transparent electrode on top. When a current applied perpendicular to the plane of these layers passes through the pinned layer, it becomes spin polarized and when this spin-polarized current is directed to the free layer, its angular momentum can be transferred to this layer, changing its magnetization direction. Reproduced with permission. [74] 2009, IEEE.

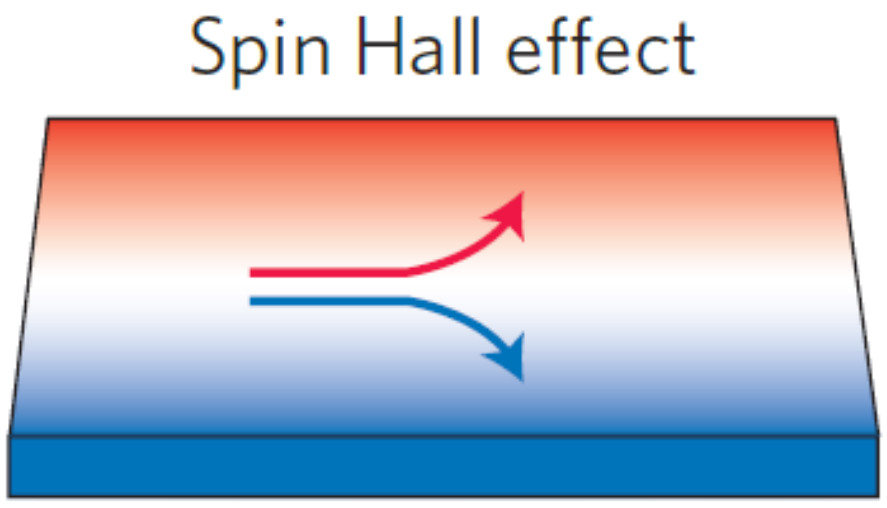

**Figure 14.** Schematic concept of Spin Hall Effect (SHE). By applying a charge current, the coupling between spin and orbital motion of electron (spin-orbit coupling) causes electrons with spin up to deflect in one direction perpendicular to the current flow and electrons with spin down in the opposite direction. As a result, an unpolarized charge current converts to a pure spin current transverse to the applied charge current. Since the number of spin up and spin down electrons are equal in an unpolarized current, a net spin flow without any charge flow will be produced in the direction perpendicular to the applied current. This spin current can apply torque on magnetic moments in a material and switch its magnetization direction. Reproduced with permission. [242] 2015, Springer Nature.

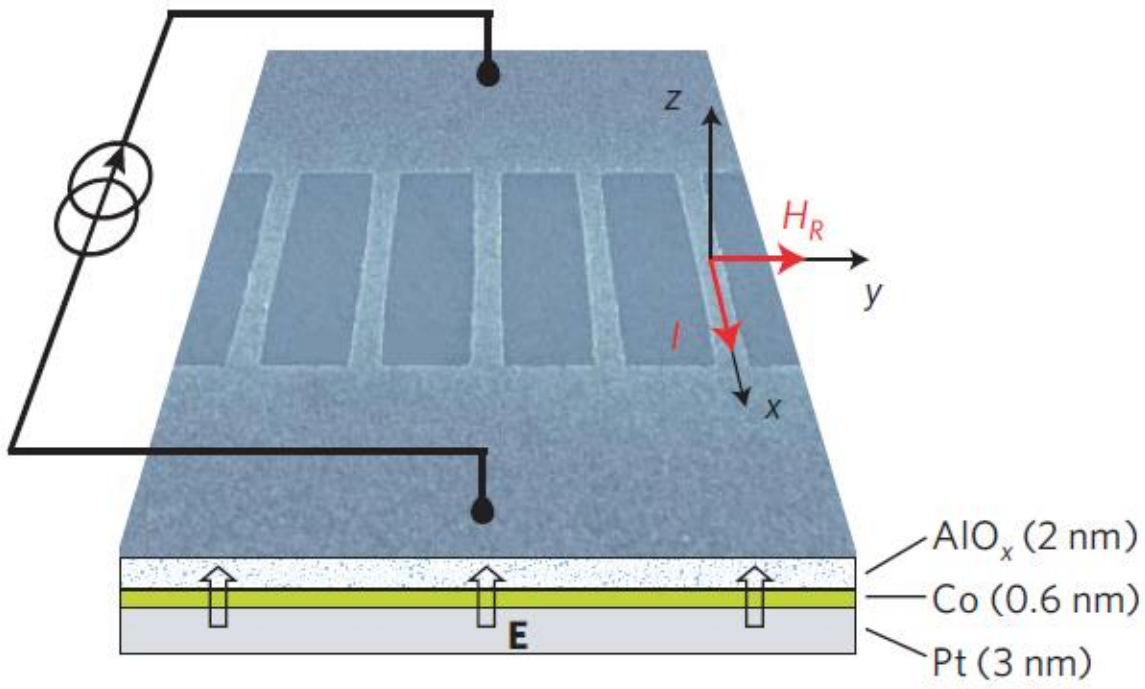

**Figure 15.** Rashba field produced by a charge current in typical heterostructures exhibiting spin-orbit torques (SOT). E shows the net electric field due to asymmetric electric potential in the direction with structural inversion asymmetry (SIA). Because of spin-orbit interactions, in the presence of a charge current, this E field transforms to an effective magnetic field $H_R$ and induces a non-equilibrium spin density perpendicular to the current direction. These spins exert

a torque on magnetic moment of the material and provoke magnetization reversal. Reproduced with permission. [244] 2010, Springer Nature.

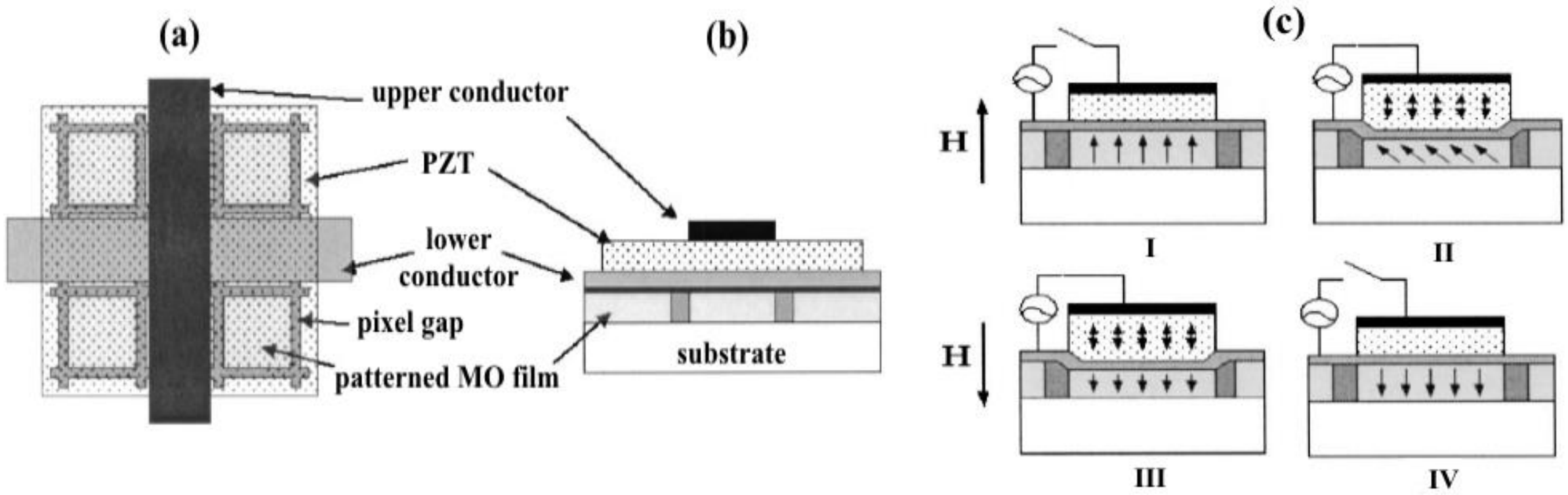

**Figure 16.** Multiferroic switching by magnetoelectric effect through strain coupling. Schematic (a) top view and (b) cross section of pixel structure for a multiferroic voltage-driven MOSLM in which switching happens through magnetoelectric effect with strain coupling. Top view represents 3×3 pixels and conductor lines are shown only on the central pixel for clarity. Pixels are comprised of a PZT layer on top of the MO garnet film. (c) Switching procedure: by applying a voltage through the crossbar metallic contacts over a pixel with out-of-plane magnetization, a stress is generated by the electrostrictive PZT layer on the selected pixel. This stress acts as an effective field which helps the magnetized domains to realign to the film plane due to magnetostriction effect. Now, a small bias field in the direction opposite to the initial magnetization causes the selected pixel to easily switch and retain that state even after turning off the voltage. Reproduced with permission. [261] 2003, AIP Publishing.

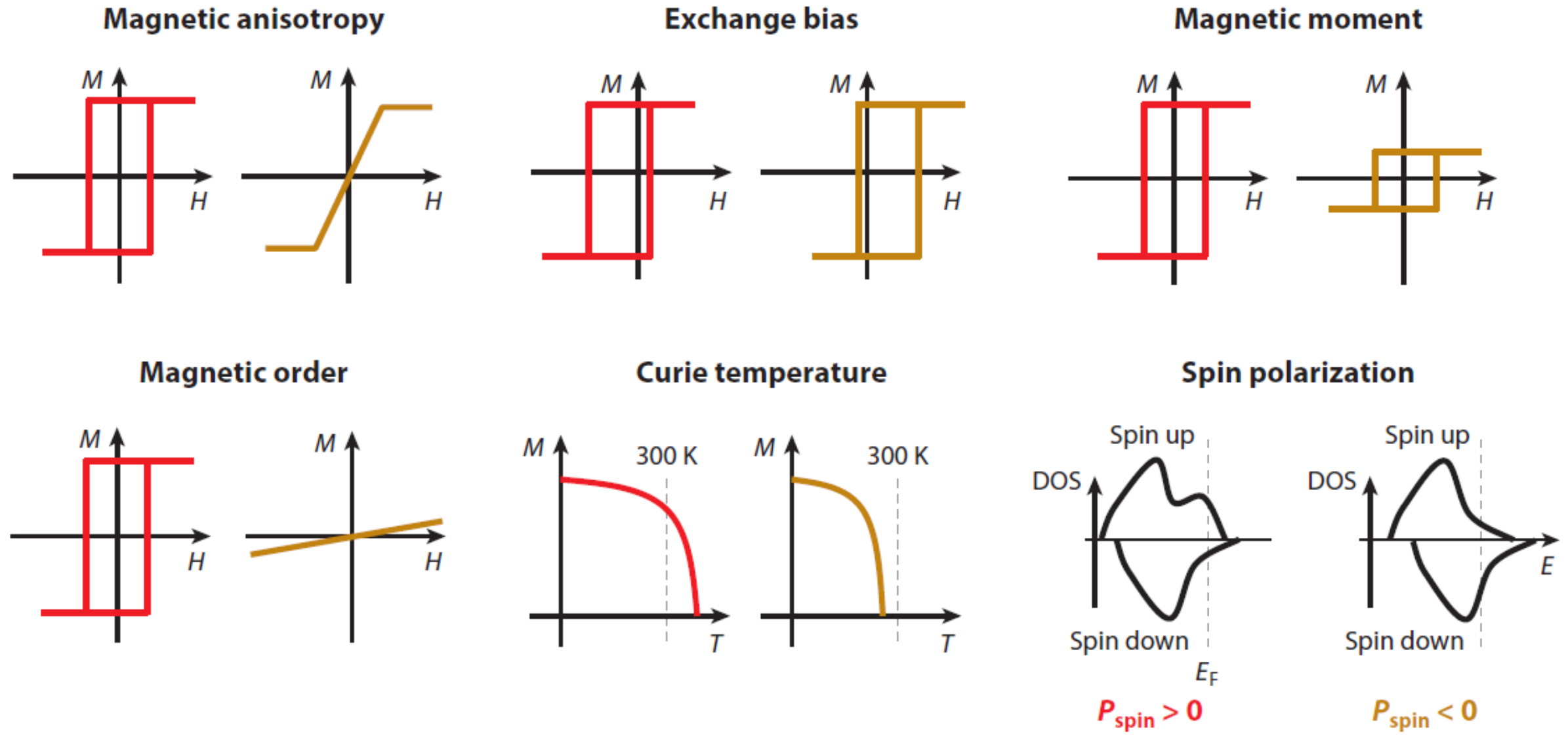

**Figure 17.** Magnetic properties that can be controlled by an electric field and therefore, can be considered as potential mechanisms for low-power magnetization switching (DOS: density of states). Reproduced with permission. [70] 2014, Annual Reviews.

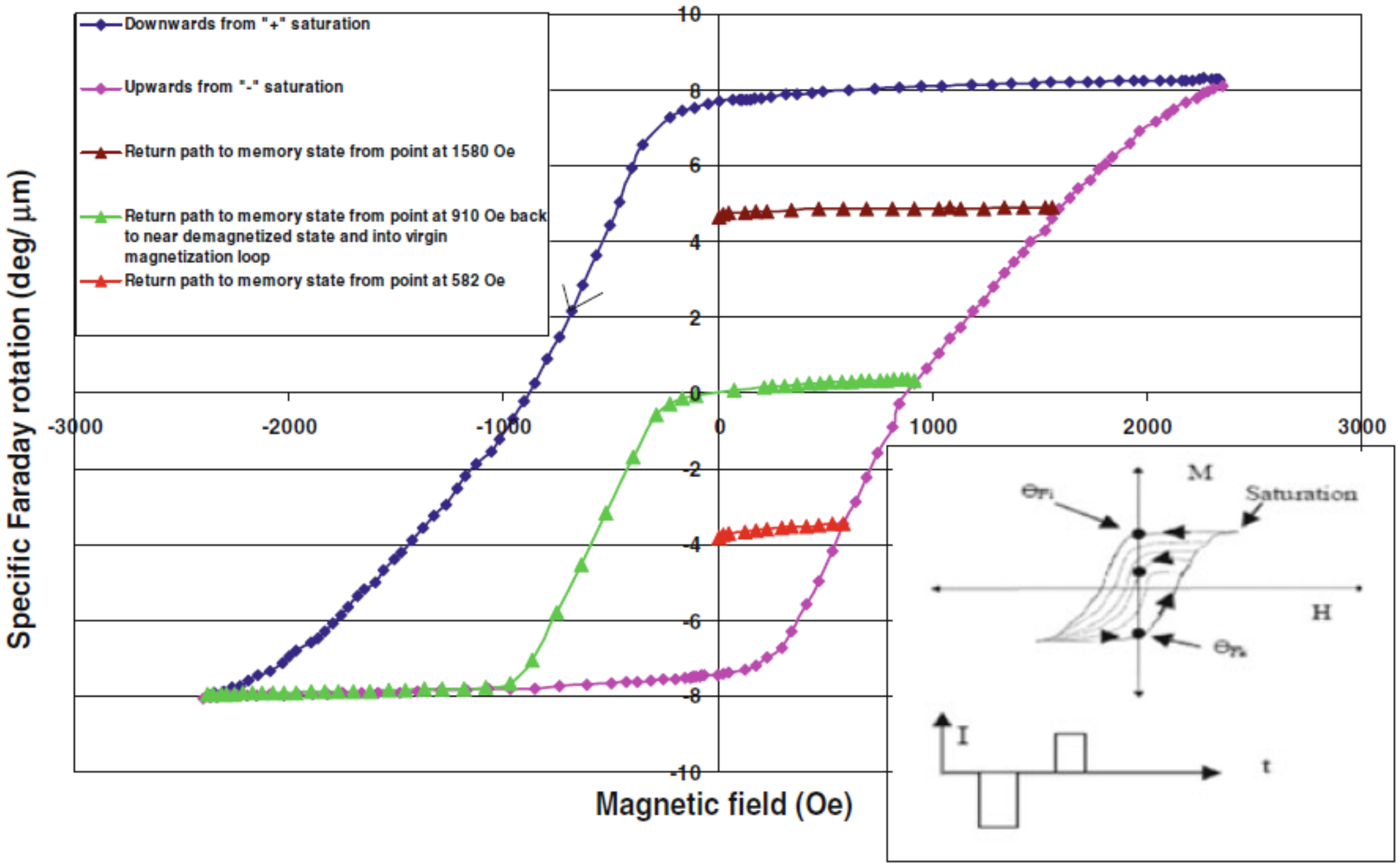

**Figure 18.** Major and minor Faraday rotation hysteresis loops in a 0.85 µm-thick garnet film grown on a Corning® 1737 glass substrate measured using 532 nm light. The minor loops show the return paths from several different magnetization states. The inset shows a schematic diagram of a magnetic hysteresis loop including some minor loops that allow for control over the remnant states of magnetization using current pulse modulation scheme. Reproduced with permission. [275] 2009, Springer.

**Table 1.** Comparison of specifications of state-of-the-art SLMs. LC, LCOS, FLCOS and DMD stand for liquid crystal, liquid crystal on silicon, ferroelectric liquid crystal on silicon, and digital micromirror device, respectively.

| Device Manufacturer Model | Type | Modulation mode | Response time /Frame rate | Pixel pitch (µm) | Resolution (pixels) | Power consumption (W) | Waveband (nm) |
|---|---|---|---|---|---|---|---|
| HOLOEYE LC 2012 [276] | LC Translucent | Phase/amplitude Analog | 60 Hz | 36 | 1024 × 768 | 1-2 | 420-800 |
| HOLOEYE GAEA-2 [277] | LCOS Reflective | Phase Analog | 58 Hz | 3.74 | 4160 × 2464 | 12 | 420-1100 1400-1700 |
| Jasper Display Corp. SRK4K – JD7714 [278] | LCOS Reflective | Phase Analog | 30 Hz | 3.74 | 4096 × 2400 | 24 | 430-750 |
| Hamamatsu X13138 series [279] | LCOS Reflective | Phase Analog | 60 Hz | 12.5 | 1280 × 1024 | 50 | 400-1550 |
| Meadowlark Optics ODP512 [280] | LCOS Reflective | Phase Analog | 3-6 ms | 15 | 512 × 512 | 15 | 400-1650 |
| Forth Dimension Displays M180 [281] | FLCOS Reflective | Phase/amplitude Binary | 75 Hz | 8.2 | 2048 × 2048 | - | 430-700 |
| Texas Instruments DLP 4710 [282] | DMD Reflective | Amplitude Binary | 10 µs | 5.4 | 1920 × 1080 | 0.804 | - |

**Table 2.** MOSLM prototype demonstrations.

| Ref., year | Modulation mode | Switching time /Frame rate | Pixel pitch (μm) | Number of pixels | Switching requirements | Wavelength (nm) |
|---|---|---|---|---|---|---|
| Krumme et al. [156], 1977 | Amplitude Binary | 10 μs | 20 | 384 × 384 | 1 μW + 100 Oe + $1.2\times10^4$ V·cm$^{-1}$ | 632.8 |
| Ross et al. [20], 1983 | Amplitude Binary | 1 μs | 100 | 48 × 48 | 1 W | - |
| Cho et al. [65], 1994 | Amplitude Binary | > 1 kHz | 24 | 128 × 128 | > 100 mW | 685 |
| Park et al. [261], 2003 | Amplitude Binary | - | 105 | 5 × 5 | 40 Oe + 8 V (sinosidal, 10 MHz) | - |
| Park et al. [38], 2003 | Amplitude Binary | - | 18 | 16 × 16 | 45 Oe (or 16 mA) | 532 |
| Park et al. [226], 2004 | Amplitude Binary | - | 18 | 16 × 16 | 120 mA + 20 Oe | - |
| Iwasaki et al. [51], 2006 | Amplitude Binary | 25 ns | 16 | 128 × 128 | 879 mW | 532 |

**Table 3.** Figure of merit (FoM) values calculated for different magnetooptical (MO) materials based on the values reported in the literature. SFR, α, $\varepsilon_{xx}$, $\varepsilon_{xy}$, FR, and k stand for specific Faraday rotation, absorption coefficient, diagonal element of permittivity tensor, non-diagonal element of permittivity tensor, Faraday rotation, and imaginary part of complex refractive index, respectively.

| Material | Synthesis method | Blue (474 nm) | | Green (520 nm) | | Red (633 nm) | | Ref. |
|---|---|---|---|---|---|---|---|---|
| | | Reported values | Calculated FoM | Reported values | Calculated FoM | Reported values | Calculated FoM | |
| **$Y_3Fe_5O_{12}$ (T<20K)** | Epitaxial flux growth | SFR = 1.45 °·μm$^{-1}$ α= 15333 cm$^{-1}$ | 0.22 °·dB$^{-1}$ | SFR = 0.23 °·μm$^{-1}$ α = 645 cm$^{-1}$ | 0.82 °·dB$^{-1}$ | SFR = 0.04 °·μm$^{-1}$ α = 245 cm$^{-1}$ | 0.38 °·dB$^{-1}$ | [283] [284] [97] |
| **$Bi_{0.7}Y_{2.3}F_5O_{12}$** | Flux technique | - | - | FoM= 2.75 °·dB$^{-1}$ | 2.75 °·dB$^{-1}$ | FoM = 3.22 °·dB$^{-1}$ | 3.22 °·dB$^{-1}$ | [285] |
| **$Y_3Fe_5O_{12}$** | Milling and sintering | $\varepsilon_{xx}$ = 5.94 +0.15060i $\varepsilon_{xy}$ = 0.00801 -0.00251i | 0.24 °·dB$^{-1}$ | $\varepsilon_{xx}$ = 5.52 +0.02410i $\varepsilon_{xy}$ = 0.00036 +0.00012i | 0.09 °·dB$^{-1}$ | $\varepsilon_{xx}$ = 5.34 +0i $\varepsilon_{xy}$ = 0.00084 +0.00084i | 0.08 °·dB$^{-1}$ | [91] |
| **$Bi_1Y_2Fe_5O_{12}$** | Milling and sintering | $\varepsilon_{xx}$ = 7.19 +0.55723i $\varepsilon_{xy}$ = -0.06631 -0.14457i | 1.45 °·dB$^{-1}$ | $\varepsilon_{xx}$ = 6.64 +0.03313i $\varepsilon_{xy}$ = -0.07499 -0.00634i | 9.62 °·dB$^{-1}$ | $\varepsilon_{xx}$ = 6.04 +0i $\varepsilon_{xy}$ = -0.01697 +0.00176i | 1.18 °·dB$^{-1}$ | [91] |
| **$FeBO_3$** | Vapor-phase transport | SFR = 0.42 °·μm$^{-1}$ α = 418 cm$^{-1}$ | 2.31 °·dB$^{-1}$ | SFR = 0.21 °·μm$^{-1}$ α = 21 cm$^{-1}$ | 23.03 °·dB$^{-1}$ | SFR = 0.08 °·μm$^{-1}$ α = 98 cm$^{-1}$ | 1.88 °·dB$^{-1}$ | [97] [286] |
| **$NiFe_2O_4$** | Conventional ceramic technology | n = 2.53 +0.79059i $\varepsilon_{xy}$ = 0.01326 -0.01889i | 0.04 °·dB$^{-1}$ | n = 2.62 +0.68830i $\varepsilon_{xy}$ = -0.00072 -0.00923i | 0.02 °·dB$^{-1}$ | n = 2.58 +0.38044i $\varepsilon_{xy}$ = -0.00077 -0.00005i | 0.002 °·dB$^{-1}$ | [287] |
| **$Y_{1.93}Bi_{1.07}Fe_5O_{12}$** | Liquid phase epitaxy | $\varepsilon_{xx}$ = 7.70 +2.00718i $\varepsilon_{xy}$ = -0.02954 -0.16213i | 0.47 °·dB$^{-1}$ | $\varepsilon_{xx}$ = 7.95 +0.89612i $\varepsilon_{xy}$ = 0.06470 -0.04779i | 0.51 °·dB$^{-1}$ | $\varepsilon_{xx}$ = 6.40 +0.04003i $\varepsilon_{xy}$ = 0.01536-0.002064i | 1.97 °·dB$^{-1}$ | [288] |
| **$(BiDy)_3(FeGa)_5O_{12}$** | Sputtering | $\varepsilon_{xx}$ = 6.96 +1.290393i $\varepsilon_{xy}$ = -0.06934 -0.00559i | 0.35 °·dB$^{-1}$ | $\varepsilon_{xx}$ = 5.79 +0.51333i $\varepsilon_{xy}$ = 0.00020 +0.01680i | 0.19 °·dB$^{-1}$ | $\varepsilon_{xx}$ = 5.20 +0.09728i $\varepsilon_{xy}$ = 0.01164 -0.00233i | 0.51 °·dB$^{-1}$ | [289] |
| **$Lu_{2.5}Bi_{0.5}Fe_5O_{12}$** | Liquid phase epitaxy | $\varepsilon_{xx}$ = 6.88 +0.60002i $\varepsilon_{xy}$ = 0.00588 -0.05661i | 0.55 °·dB$^{-1}$ | $\varepsilon_{xx}$ = 6.43 +0.19270i $\varepsilon_{xy}$ = -0.01176 +0.01342i | 0.41 °·dB$^{-1}$ | $\varepsilon_{xx}$ = 5.78 +0.02631i $\varepsilon_{xy}$ = -0.00323 +0.00506i | 0.02 °·dB$^{-1}$ | [93] |
| **$Lu_{2.3}Bi_{0.7}Fe_{4.4}Ga_{0.6}O_{12}$** | Liquid phase epitaxy | $\varepsilon_{xx}$ = 6.48 +0.44424i $\varepsilon_{xy}$ = -0.00323 -0.07393i | 0.90 °·dB$^{-1}$ | $\varepsilon_{xx}$ = 6.17 +0.16254i $\varepsilon_{xy}$ = -0.02049 -0.01245i | 0.49 °·dB$^{-1}$ | $\varepsilon_{xx}$ = 5.62 +0.02133i $\varepsilon_{xy}$ = -0.00607 +0.00370i | 0.21 °·dB$^{-1}$ | [93] |
| **$Bi_3Fe_5O_{12}$** | Rf-magnetron sputtering | - | - | FR = 11.83 ° T = 0.71 % | 0.55 °·dB$^{-1}$ | FR = 5.38 ° T = 69 % | 3.34 °·dB$^{-1}$ | [67] |
| **$Gd_{1.24}Pr_{0.48}Bi_{1.01}Lu_{0.27}Fe_{4.38}Al_{0.6}O_{12}$** | Liquid phase epitaxy | $\varepsilon_{xx}$ = 7.52 +0.96644i $\varepsilon_{xy}$ = 0.13229 -0.05390i | 0.91 °·dB$^{-1}$ | $\varepsilon_{xx}$ = 6.82 +0.12387i $\varepsilon_{xy}$ = 0.01803 -0.05864i | 2.88 °·dB$^{-1}$ | $\varepsilon_{xx}$ = 6.06 +0.03278i $\varepsilon_{xy}$ = -0.00130 -0.01759i | 1.15 °·dB$^{-1}$ | [290] |
| **$Ce_1Y_2Fe_5O_{12}$** | Pulsed laser deposition | FoM = 0.12 °·dB$^{-1}$ | 0.12 °·dB$^{-1}$ | FoM = 0.096 °·dB$^{-1}$ | 0.096 °·dB$^{-1}$ | FoM = 0.16 °·dB$^{-1}$ | 0.16 °·dB$^{-1}$ | [291] |
| **$Bi_{1.5}Y_{1.5}Fe_5O_{12}$** | Metal organic decomposition | SFR= 5.82 °·µm$^{-1}$ k= 0.2869 | 0.18 °·dB$^{-1}$ | SFR = 8.16 °·µm$^{-1}$ k = 0.1913 | 0.41 °·dB$^{-1}$ | SFR = 1.90 °·µm$^{-1}$ k = 0.0171 | 1.29 °·dB$^{-1}$ | [96] |
| **$Bi_2Y_1Fe_5O_{12}$** | Metal organic decomposition | SFR= 14.60 °·µm$^{-1}$ k= 0.3263 | 0.39 °·dB$^{-1}$ | SFR = 16.05 °·µm$^{-1}$ k = 0.2093 | 0.73 °·dB$^{-1}$ | SFR = 3.47 °·µm$^{-1}$ k = 0.0168 | 2.40 °·dB$^{-1}$ | [96] |
| **$Bi_{2.5}Y_{0.5}Fe_5O_{12}$** | Metal organic decomposition | SFR= 24.06 °·µm$^{-1}$ k= 0.4477 | 0.47 °·dB$^{-1}$ | SFR = 20.1 °·µm$^{-1}$ k = 0.2450 | 0.78 °·dB$^{-1}$ | SFR = 5.72 °·µm$^{-1}$ k = 0.0134 | 4.95 °·dB$^{-1}$ | [96] |
| **$Bi_3Fe_5O_{12}$** | Metal organic decomposition | SFR= 35.27 °·µm$^{-1}$ k= 0.5327 | 0.58 °·dB$^{-1}$ | SFR = 25.44 °·µm$^{-1}$ k = 0.2912 | 0.83 °·dB$^{-1}$ | SFR = 6.84 °·µm$^{-1}$ k = 0.0069 | 11.49 °·dB$^{-1}$ | [96] |

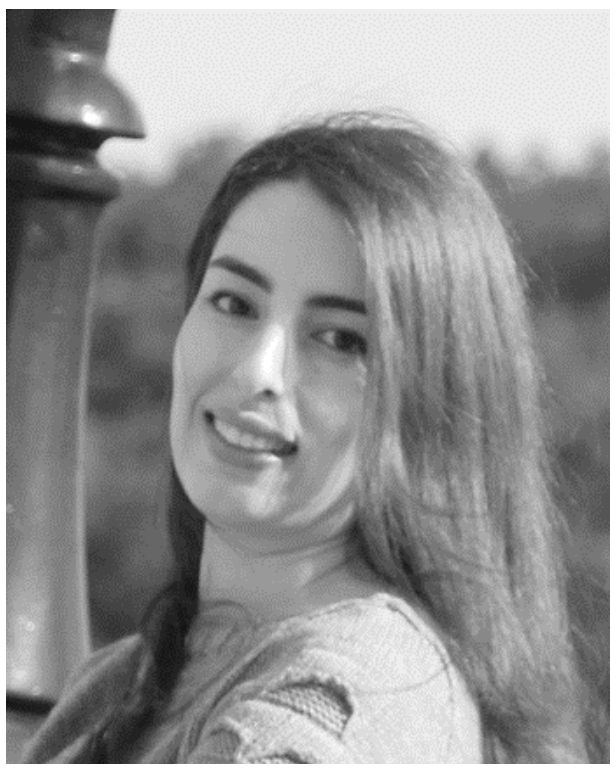

Soheila Kharratian received her B.Sc. degree in Materials Science and Engineering from University of Tabriz, Tabriz, Iran in 2010 and her M.Sc. degree from University of Tehran, Tehran, Iran in 2014. Her Master's thesis included electrodeposition and characterization of magnetic soft-hard composite thin films for MEMS applications. She joined Koç University as a Ph.D. student in Materials Science and Engineering in 2015. She spent about three months as a Visiting Researcher at the École Polytechnique Fédérale de Lausanne, Lausanne, Switzerland. She is currently working on magnetooptical materials and spatial light modulators. Her research interests include magnetooptics, photonics, plasmonics and nanostructured materials.

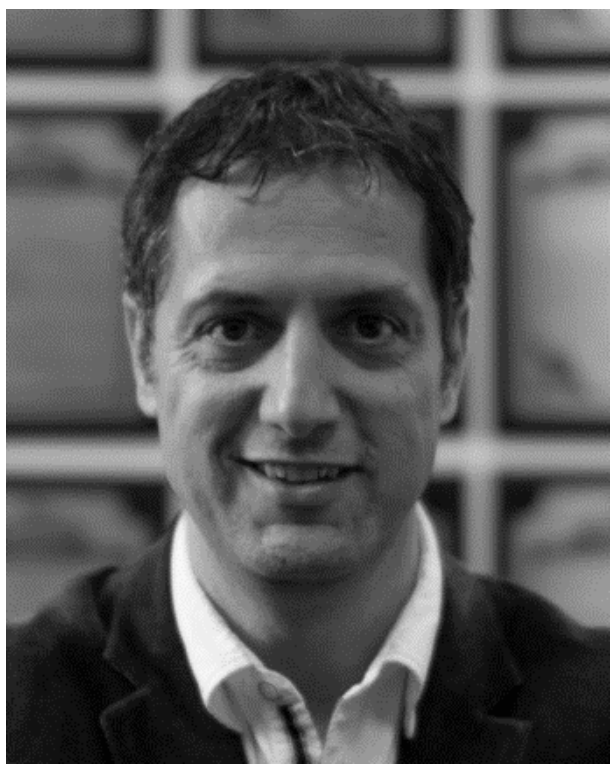

Hakan Urey received the B.Sc. degree from Middle East Technical University, Ankara in 1992, and M.Sc. and Ph.D. degrees from Georgia Institute of Technology in 1996 and 1997, respectively, all in Electrical Engineering. After completing his Ph.D., he worked on Retinal Scanning Display technology in Microvision lnc. He is currently a professor in Electrical and Electronics Department at Koç University. His research interests are in the area of optical

MEMS, micro-optics and optical system design, 2D/3D display and imaging systems, and biosensors. He has published more than 180 journal and conference papers, edited books and book chapters, and has more than 50 issued and several pending patents. He has received many outstanding grants and awards in the field, including an advanced grant from the European Research Council (ERC-AdG) in 2013.

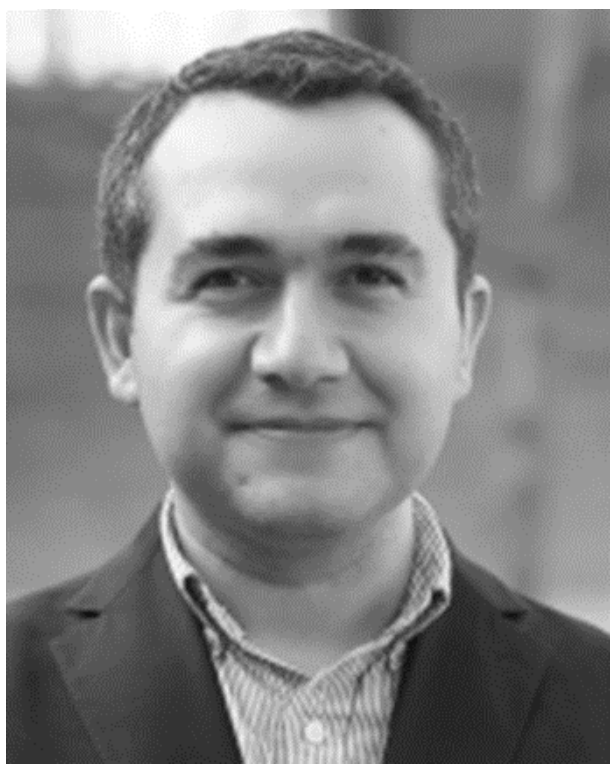

Mehmet C. Onbaşlı is currently an assistant professor at the Department of Electrical and Electronics Engineering at Koç University. He received his PhD from Massachusetts Institute of Technology in 2015 from Materials Science and Engineering Department and his B.Sc. degree from Electrical and Electronics Engineering Department, Bilkent University, in 2010. During his PhD, he worked on the growth, characterization, device fabrication and testing of magnetic, magnetooptic and ferroelectric thin films. After working at Corning Inc. for one year on data-driven glass and ceramics modeling, he joined Koç University. His research interests include magnetooptics, magnetic memory elements, spintronic, logic circuits, sensors and their applications in biological systems. He has published over 80 journal or conference papers and has a granted patent on magnetooptical isolators and garnet processing.